# Cell-penetrating pepducins targeting the neurotensin receptor type 1 relieve pain

Rebecca L. Brouillette[1‡], Élie Besserer-Offroy[2‡*], Christine E. Mona[3], Magali Chartier[1], Sandrine Lavenus[1], Marc Sousbie[1], Karine Belleville[1], Jean-Michel Longpré[1], Éric Marsault[1], Michel Grandbois[1], Philippe Sarret[1*].

[1]Department of Pharmacology-Physiology, Faculty of Medicine and Health Sciences, Institut de pharmacologie de Sherbrooke, Université de Sherbrooke, Sherbrooke, QC, Canada

[2]Department of Pharmacology and Therapeutics, McGill University, Montreal, QC, Canada

[3]Ahmanson Translational Theranostic Division, Department of Molecular and Medical Pharmacology, David Geffen School of Medicine, University of California at Los Angeles, Los Angeles, CA, USA

[‡] **Equal contribution**

[*]**Corresponding authors**

**Philippe Sarret, Ph.D.**
Dept. of Pharmacology-Physiology
Faculty of Medicine and Health Sciences
Université de Sherbrooke
3001, 12th Avenue North
Sherbrooke, Québec, J1H 5H4
Canada
Tel: (819) 821-8000, Ext: 72554
Philippe.Sarret@USherbrooke.ca

**Élie Besserer-Offroy, Ph.D.**
Dept. of Pharmacology and Therapeutics
McGill University
McIntyre Medical Sciences Building
3655 Sir William Osler Promenade
Montréal, Québec, H3G 1Y6
Canada
Tel: (514) 393-8803
Elie.Besserer@McGill.ca

**Abstract**

Pepducins are cell-penetrating, membrane-tethered lipopeptides designed to target the intracellular region of a G protein-coupled receptor (GPCR) in order to allosterically modulate the receptor's signaling output. In this proof-of-concept study, we explored the pain-relief potential of a pepducin series derived from the first intracellular loop of neurotensin receptor type 1 (NTS1), a class A GPCR that mediates many of the effects of the neurotensin (NT) tridecapeptide, including hypothermia, hypotension and analgesia. We used BRET-based biosensors to determine the pepducins' ability to engage G protein signaling pathways associated with NTS1 activation. We observed partial $G\alpha_q$ and $G\alpha_{13}$ activation at a 10 μM concentration, indicating that these pepducins may act as allosteric agonists of NTS1. Additionally, we used surface plasmon resonance (SPR) as a label-free assay to monitor pepducin-induced responses in CHO-K1 cells stably expressing hNTS1. This whole-cell integrated assay enabled us to subdivide our pepducin series into three profile response groups. In order to determine the pepducins' antinociceptive potential, we then screened the series in an acute pain model (tail-flick test) by measuring tail withdrawal latencies to a thermal nociceptive stimulus, following intrathecal (i.t.) pepducin administration (275 nmol/kg). We further evaluated promising pepducins in a tonic pain model (formalin test), as well as in neuropathic (*Chronic Constriction Injury*) and inflammatory (*Complete Freund's Adjuvant*) chronic pain models. We report one pepducin, PP-001, that consistently reduced rat nociceptive behaviors, even in chronic pain paradigms. Finally, we designed a TAMRA-tagged version of PP-001 and found by confocal microscopy that the pepducin reached the rat dorsal root ganglia post i.t. injection, thus potentially modulating the activity of NTS1 at this location to produce its analgesic effect. Altogether, these results suggest that NTS1-derived pepducins may represent a promising strategy in pain-relief.

**Key words:** pepducins, lipopeptides, neurotensin, pain, surface plasmon resonance, drug discovery, G protein-coupled receptor (GPCR).

## 1. Introduction

G protein-coupled receptors (GPCRs), alternatively known as seven transmembrane receptors (7TMRs), represent the single most successful protein target class in drug discovery. Typically, 30 – 50% of all approved drugs on the market today modulate this family of cell surface receptors (Overington, et al., 2006). In recent years, a number of new pharmacological approaches have emerged to further exploit their therapeutic potential. One such strategy is the use of cell-penetrating, membrane-tethered lipopeptides called pepducins (Covic, et al., 2002).

Specifically, pepducins are composed of a peptide sequence derived from one of the intracellular loops of a target 7TM receptor to which a lipid moiety, most commonly a palmitic acid, is conjugated at the N-terminal end. Here, the palmitate fulfills the role of a tether which effectively anchors the peptide to the cell membrane and enables its translocation toward the cytosolic phospholipid leaflet, where it acts at the intracellular receptor-effector interface to modulate receptor signaling. Pepducins can act as allosteric agonists or allosteric modulators of their cognate receptors, potentially accessing unique receptor conformations that are unavailable with traditional orthosteric ligands (Carr III & Benovic, 2016). Since its first introduction by Covic et al. in 2002, pepducin technology has been applied to the study of more than a dozen 7TMRs in a variety of disease paradigms, including cardiovascular disease, cancer, angiogenesis, inflammation, sepsis and asthma (Zhang, et al., 2015a; Tressel, et al., 2011). Notably, a protease-activated receptor 1 (PAR1)-derived pepducin called PZ-128 will soon complete a phase 2 clinical trial as a novel antiplatelet agent, indicating that pepducins may constitute promising therapeutics (Gurbel, et al., 2016).

Nevertheless, relatively little is known about the pepducin mode of action, and there is a need to further study pepducin actions in new receptor systems and physiological conditions. In particular, the potential applications of the pepducin approach to pain treatment remain largely unexplored. To date, very few studies have administered pepducins to rodents in a pain context, and these center almost exclusively around a PAR4-derived pepducin (P4pal-10), used as a tool to decipher the role of the PAR4 receptor in pain and inflammation (McDougall, et al., 2009; Russell, et al., 2009; Annahazi, et al., 2009). To our knowledge, this study is the first to evaluate pepducins in a chronic pain paradigm.

From the perspective of pain-relief, neurotensin receptor type 1 (NTS1) is a particularly interesting GPCR target candidate. NTS1 belongs to the rhodopsin-like family of 7TMRs (Class A) (Tanaka, et al., 1990; Vita et al., 1993) and is expressed peripherally in the gut (Azriel & Burcher, 2001), in certain cardiac and endothelial cells (Tanaka, et al., 1990; Schaeffer, et al., 1995), and in pain-related regions of the central nervous system, such as the periaqueductal gray, the superficial laminae of the spinal cord dorsal horn, and the dorsal root ganglia (DRGs) (Roussy et al., 2008). Consistent with its expression pattern, NTS1 activation has been linked to a wide range of physiological effects, such as hypotension (Rioux, et al., 1982), ileum contraction and relaxation (Carraway and Mitra, 1994), hypothermia (Feifel, et al., 2010), and, notably, analgesia (Pettibone, et al., 2002; Roussy, et al., 2008). Endogenously, it is activated by the neurotensin (NT) tridecapeptide (Carraway & Leeman, 1973), which also mediates its effects via another 7TMR (NTS2) and a Type 1 transmembrane monomeric glycoprotein (NTS3/sortilin) (Mazella & Vincent, 2006). When administered centrally to rodents, NT is known to produce a very potent analgesic action. Indeed, on a molar basis, NT is more potent than morphine at producing antinociception (Nemeroff, et al., 1979). Furthermore, this antinociceptive effect appears to be independent of the opioid system, as the opioid receptor antagonists naloxone and naltrexone have no effect on NT-mediated antinociception (Clineschmidt & McGuffin, 1977). Furthermore, our research team has shown that NTS1 agonists can effectively reverse nociceptive behaviors in rat models of acute and tonic pain (Roussy, et al., 2008), as well as in chronic neuropathic pain models (Guillemette, et al., 2012). Until now, no pepducins have been developed to target NTS1.

Herein, we provide proof-of-concept data highlighting the potential of the pepducin approach in the field of pain research. For this study, we designed and synthesized a pepducin series derived from the first intracellular loop of NTS1, which we then screened in BRET-based and label-free cellular assays monitoring G protein activation and whole-cell integrated responses, respectively. We further evaluated the antinociceptive potential of these pepducins for different types of pain, and hereby report a pepducin that consistently reduced rat nociceptive behaviors, when injected intrathecally (i.t.).

## 2. Materials and Methods

### *2.1 Materials*

In order to perform the experiments described here, we obtained the NTS1 antagonist SR48692 from R & D Systems, Inc. (Minneapolis, MN, USA). We synthesized the compounds NT(8-13), and our NTS1-derived pepducin series at the peptide synthesis facility of the Université de Sherbrooke (http://www.usherbrooke.ca/ips/en/platforms/psp/). This series consists of five main pepducins (NTS1-ICL1-PP-001 to NTS1-ICL1-PP-005), a non-palmitoylated control (NTS1-ICL1-NP-001), a pepducin with a scrambled peptide sequence (NTS1-ICL1-PP-SCR-001), as well as a TAMRA-labeled pepducin (NTS1-ICL1-FP-001). The specific peptide sequences are provided in **Table 1**. The protected amino acids and TentaGel R-RAM resins used in their synthesis were obtained from ChemImpex International (Wood Dale, IL, USA). All other chemicals and reagents involved in peptide and pepducin synthesis were obtained from Sigma-Aldrich (Oakville, ON, Canada) or Fisher Scientific (Montreal, QC, Canada). DMEM-F12, HEPES (4-(2-hydroxyethyl)-1-piperazineethanesulfonic acid), penicillin-streptomycin-glutamine, fetal bovine serum (FBS), gentamycin G418, phosphate-buffered solution (PBS), Hank's balanced salt solution (HBSS), trypsin-EDTA 0.25% were all obtained from Wisent (St. Bruno, QC, Canada).

### *2.2 Pepducin synthesis*

The ICL1-derived pepducin series was synthesized using standard C→N solid-phase peptide synthesis. The peptide chains for each of the pepducins in this series were synthesized on a 0.1 mmol scale using manual solid phase peptide synthesis. Peptide synthesis was performed in 6 mL polypropylene cartridge with 20 μm PE frit from Applied Separations (Allentown, PA, USA). The 6 mL reactors were shaken (~180 rpm) on New Brunswick Scientific orbital shaker. To perform these syntheses, Rink Amide resins (Chem-Impex, Wood Dale, IL, USA) with a loading capacity of 0.51 meq/g was used. As a first step, the Fmoc-protecting group from the Rink Amide resin was removed using 50% piperidine in DMF for 2 cycles of 10 minutes. The solvent was subsequently filtered, and the resin was washed with alternative 2-propanol and dichloromethane (DCM, 5 mL) cycles, and 3 final cycles of consecutive DMF (3x8 mL). Kaiser test was performed to ensure removal of the Fmoc group then the deprotected resin was reacted with the first Fmoc N-protected amino acid (5 eq.) in the presence of HATU (5 eq.), N,N-diisopropylethylamine (10 eq.) in DMF (2 mL) for at least 2 hours. The resin was washed after each coupling and deprotection step by shaking 5 min in DMF (2 × 5 mL), then with three alternative washing cycles of 2-propanol (5 mL) or DCM (5 mL). The rest of the synthesis for the peptide elongation was performed as indicated above. For difficult coupling, we used microwave-assisted synthesis and COMU for more efficient peptide coupling. The completion of the coupling reaction was assessed using the Kaiser test.

In the case of the non-palmitoylated control, the peptide was cleaved from the resin upon sequence completion, following removal of the last Fmoc protecting group via standard deprotection conditions and washes. In the case of the pepducins, the palmitic acid was coupled to the free N-terminal end of the peptide sequence. The resin was again submitted to the Fmoc de-protection steps and dried with diethyl ether. Palmitic acid (5 eq.) and HATU (5 eq.) were dissolved in an anhydrous NMP/DCM (1:1) mix followed by DIPEA (10 eq.). The coupling mixture was then added to the reaction vessel and shaken overnight at room temperature. The resin was washed extensively after coupling with three alternate cycles of 2-propanol (5 mL) or DCM (5 mL). Final resin wash was done in diethyl ether prior to cleavage. For the synthesis of the TAMRA fluorescent pepducin (NTS1-ICL1-FP-001), 2-(((9H-fluoren-9-yl)methoxy)carbonylamino)octadecanoic acid was purchased from Sigma-Aldrich (Oakville, ON, Canada), and was incorporated to the peptide as an amino acid using procedures previously described. Following the deprotection of the Fmoc, using the standard deprotection protocol, the fluorescent probe 5(6)- Carboxytetramethylrhodamine (5(6)-TAMRA, 4 eq., EMD) was coupled overnight to the free amine using the same approach described for palmitic acid.

Peptides were cleaved using 95% TFA, 2.5% TIPS, 2.5% $H_2O$ for at least 2 hours to provide the crude pepducins. The resin suspension was filtered through cotton wool and the peptide precipitated in a 50 mL Falcon tube in tert-butyl methyl ether (TBDME) at 0 °C. The suspension was centrifuged to a pellet for 20 minutes at 1500 rpm and the filtrate was removed by decantation. The pellet was placed under high vacuum to remove residual traces of solvent, then resuspended in water/acetonitrile and lyophilized to a powder. A sample of the crude pepducin was analyzed by UPLC/MS (Waters UPLC system coupled with a SQ detector 2 and a PDA eλ detector, coupled to an Acquity UPLC BEH C18 column, 2.1 mm X 50 mm, 1.7 µm spherical size). If the desired mass was displayed by the UPLC/MS, usually represented as [M+2]/2 or [M+3]/3, the compound was purified by MS-triggered HPLC (Autosampler 2707, quaternary gradient module 2535, UV detector 2489, fraction collector WFCIII, from Waters (Milford, MA, USA)), using a C18 column (250 mm X 21.2 mm, 5 µm spherical size). UPLC chromatograms were recorded on a Waters Acquity H class using the following gradient: water + 0.1% TFA and acetonitrile (0 → 0.2 min, 5% acetonitrile; 0.2 → 1.5 min, 5% → 95%; 1.5 → 1.8 min, 95%; 1.8 → 2.0 min, 95% → 5%; 2.0 → 2.5 min, 5%). The crude pepducin (30-60 mg) was dissolved using sonication in 3 mL of a DMSO:ACN:$H_2O$ mix (1:1:1). The pepducins were purified using variable gradients adapted to the observed analytical retention time; fractions containing the desired mass were combined and lyophilized. The lyophilized pepducins retrieved with 95% purity were then dissolved in 100% DMSO. Stock solutions were prepared for biological assays at 10 mM concentrations. The fluorescent pepducin was kept away from light throughout all the synthesis, purification and storage process.

### ***2.3 Cellular assays***

#### *2.3.1 Cell culture and transfections*

The biological assays used to characterize the *in vitro* behavior of our pepducin series were performed with Chinese Hamster Ovarian (CHO-K1, CCL-61 from ATCC, Manassas, VA) cells. The cells were cultured in a DMEM-F12 medium containing 20 mM HEPES, 10 % FBS, and penicillin (100 U/mL)-streptomycin (100 μg/mL)-glutamine (2 mM). They were maintained at 37$^{o}$C and 5% $CO_2$, in a humidified atmosphere. CHO-K1 cells stably expressing hNTS1, purchased from Perkin Elmer (Montreal, QC, Canada), were also cultured in the same conditions as above, but further supplemented with 0.4 mg/mL of the geneticin (G418) antibiotic. All cell lines were used between passages 5 and 20. Certain cellular assays required the transfection of cDNA plasmids, for transient expression of recombinant proteins. The transfection procedure was as follows: 1.5 x $10^6$ cells were seeded onto 100 $mm^2$ cell culture dishes, and, 24 hours later, received a total of 12 μg cDNA, prepared in Opti-MEM serum-free media along with the transfection agent polyethylenimine (PEI) at a 3:1 ratio (PEI : DNA), as previously described (Erhardt, et al., 2006).

#### *2.3.3 G protein activation cellular assay*

The pepducins' ability to activate NTS1-associated G protein signaling pathways was assessed using biosensors based on Bioluminescence Resonance Energy Transfer (BRET) technology. BRET is defined as a transfer of energy that occurs between an enzymatic donor and a fluorescent acceptor when these two proteins are at a distance closer than 100 Å and in a favorable orientation, following activation of the donor by degradation of its substrate. The BRET-based biosensors used here were designed to directly measure the dissociation of Gα and Gγ protein subunits, and were generously provided by Dr. Michel Bouvier (Department of Biochemistry and IRIC, Université de Montréal, Montréal, QC, Canada), as a member of the CQDM team (Drs. M. Bouvier, T. Hébert, S.A. Laporte, G. Pineyro, J.-C. Tardif, E. Thorin and R. Leduc). In the assays performed in this study, 1.5 x $10^6$ CHO-K1 cells were seeded into 100 $mm^2$ cell culture dishes and transfected 24 hours later according to the procedure described above. The cells were transfected with either of the following biosensor couples: hNTS1, Gαq-RlucII, $G\beta_1$ and Gγ-$GFP^{10}$ (Breton, et al., 2010); or hNTS1, Gα13-RlucII, $G\beta_1$ and Gγ-$GFP^{10}$ (Quoyer, et al., 2013; Demeule, et al., 2014). Cells were detached 24 hours post-transfection using trypsin-EDTA 0.25% and were seeded into white opaque 96-well plates (BD Falcon, Corning, NY, USA) at a concentration of 50 000 cells/well. On the final day of the experiment, the adhered cells were washed once with 100 µL of PBS and stimulated with 1 µM of NT(8-13) or 10 µM of the pepducins, prepared in a 100 µL volume of HBSS (20 mM HEPES). The cells were then stimulated with coelenterazine 400A (5 µM), incubated at 37°C for 15 minutes, and read on a GENios Pro plate reader (Tecan, Durham, NC, USA) using a $BRET^2$ filter set (400-450 nm and 500-550 nm emission filters).

For each well, a $BRET^2$ ratio was determined by dividing the $GFP^{10}$-associated light emission by RlucII-associated light emission. The data was subsequently normalized relative to NT(8-13);

values for non-treated cells were set as 0% pathway activation, and those for cells treated with 1 μM NT(8-13) were set as 100% pathway activation.

*2.3.2 Surface plasmon resonance cellular assay*

We monitored the global response of living cells to molecular stimuli by surface plasmon resonance (SPR). As previously described in Cuerrier, et al. (2008) and in Lavenus, et al. (2018), we prepared a SPR substrate that consisted of a glass support upon which a gold layer (48 nm) was deposited on top of a chromium adhesion layer (3nm), followed by a 5 min exposure with poly-L-lysine 0.01% (Sigma, Oakville, Canada) to promote cell adhesion. CHO-K1 cells stably expressing hNTS1 (CHO-hNTS1) or CHO-K1 wild-type cells were seeded at a density of 100 000 cells/cm$^2$ onto 60 mm petri dishes containing the SPR substrate and allowed to grow for two days in a DMEM-F12 medium supplemented with HEPES, FBS and antibiotics. They were starved for one day in DMEM-F12 media (without FBS) prior to the experiment. Measurements were performed at 90% confluency, as evaluated by contrast microscopy. Once placed on the SPR apparatus, the media was replaced with Leibovitz' L15 media. Once the incident angle for optimal SPR detection was determined, a 10-min baseline was recorded to ensure that the cells reached a steady state before compound incubation. The CHO-hNTS1 cells were then stimulated with 10 μM concentrations of NT(8-13) or of the ICL1-derived pepducin series. Cells were subsequently monitored for a period of 30 minutes, and data were acquired at 1-second intervals. The results for this experiment are plotted in terms of reflectance variation units (RVU) over time, for which 1 RVU represents 0.1% variation in total reflectance and the time is measured in seconds.

***2.4 Behavioral studies***

*2.4.1 Animals, housing and habituation*

All animal procedures were approved by the Ethical and Animal Care Committee of the Université de Sherbrooke (protocol number 035-18) and were in accordance with policies and directives of the Canadian Council on Animal Care. They further comply with the ARRIVE guidelines. Adult male (250−300 g) Sprague−Dawley rats (Charles River Laboratories, St-Constant, QC, Canada) were maintained on a 12-hour light/dark cycle with free access to food and water. Prior to the behavioral studies, the rats were acclimatized for 4 consecutive days to the animal facility and for 3 consecutive days to the experimental conditions of each test.

*2.4.2 I.t. administration of NTS1-derived pepducins*

In all the behavioral tests reported here, the compounds were administered via i.t. injection. Rats were lightly anesthetized with 2.5% isofluorane (Abbott Laboratories, Montreal, QC, Canada). Subsequently, a 25 µL volume of ICL1-derived pepducins at a 275 nmol/kg was administered alone or in conjunction with the SR48692 antagonist (10 µg/kg), by injection into the subarachnoid space between lumbar vertebrae L5 and L6, using a 27 G 1/2 needle. Pepducins were diluted in a vehicle composed of physiological saline, 10% DMSO and 20% polyethylene glycol 4000 (PEG4000). Control animals were injected with the vehicle alone. At these doses, no sedation or visible side effects were observed.

*2.4.3 Acute pain model (tail-flick test)*

The effect of the ICL1-derived pepducins on acute thermal nociception was assessed using the tail-flick test. This test involves measuring the latency (in seconds) for a rat to withdraw its tail from an acute nociceptive stimulus. Here, this stimulus was a water bath maintained at 52°C. The effects of the pepducins' and of the vehicle were assessed immediately following i.t. injection. The thermal threshold latencies were determined every 10 min for up to 1 hour after drug injection. A cut-off was set at 10 s to avoid tissue damage.

*2.4.4 Tonic pain model (formalin test)*

Antinociception was also assessed using the formalin test as a model of tonic pain. Following i.t. injection of the vehicle or test compounds, acclimated male Sprague-Dawley rats received a 50 μL subcutaneous injection of diluted 2% formaldehyde (i.e., 5% formalin, Fisher Scientific, Montreal, QC, Canada) into the plantar surface of the right hind paw. The rats were then placed in clear plastic chambers (30 cm × 30 cm × 30 cm) positioned over a mirror angled at 45° in order to allow an unobstructed view of the paws. The rats' behaviors were observed for a 60-minute period. The intra-plantar injection of formalin produced the biphasic nociceptive response typical of this tonic pain model. The two distinct phases of spontaneous pain behaviors that occur in rodents are proposed to reflect a direct effect of formalin on sensory receptors (acute phase; 0-9 min post-injection) and a longer-lasting pain due to inflammation and central sensitization (inflammatory phase; 21-60 min post-injection).

Nocifensive behaviors were assessed using a weighted score method, as previously described (Coderre, et al., 1993). Following the injection of formalin into the right hind paw, the experimenter measured the time spent in each of four behavioral categories: 0, the injected paw is comparable to the contralateral paw; 1, the injected paw has little or no weight placed on it; 2, the injected paw is elevated and is not in contact with any surface; 3, the injected paw is licked, bitten, or flinched. The behaviors believed to represent higher levels of pain intensity were given higher weighted scores. The weighted average pain intensity score ranging from 0 to 3 was then calculated by multiplying the time spent in each category by the category weight, summing these products, and dividing by the total time in a given time interval. The pain score was thus calculated from the following formula (1T1 + 2T2 + 3T3)/180 where T1, T2, and T3 are the duration (in seconds) spent in behavioral categories 1, 2, or 3, respectively, during each 180 s block (Dubuisson & Dennis, 1977). The total area under the curve (AUC) for the inflammatory phase was calculated between 21−60 min.

*2.4.5 Chronic neuropathic pain model (chronic constriction injury model)*

The pepducins' analgesic potential was further assessed in chronic pain models. Neuropathic pain was induced through chronic constriction injury (CCI) of the sciatic nerve, as previously described (Bennett & Xie, 1988). Briefly, under isoflurane anaesthesia (5% in $O_2$, Abbott Laboratories, Montreal, QC, Canada), the rat was placed in left lateral recumbency, and a 2.5 cm longitudinal skin incision was made starting at the sciatic notch. The right common sciatic nerve was exposed at the midthigh level by blunt dissection through the biceps femoris muscle. Proximal to the sciatic nerve's trifurcation, 7 mm of nerve was freed from adhering tissues, and four ligatures were loosely tied around it with about 1 mm of spacing (4.0 Sofsilk suture). The wound was then closed in layers and the skin was sealed with 4.0 silk sutures. This right side was defined as the operated-ipsilateral limb, while the non-operated leg (left) represented the contralateral side. Great care was taken to tie the ligatures such as the diameter of the sciatic nerve was seen to be just barely constricted, when viewed under a dissecting microscope. Rectal temperature was maintained at 37 ± 1°C by using a heating pad and ophthalmic ointment was used to prevent corneal desiccation. Rats were housed individually for 48 h following surgery.

Subsequently, mechanical allodynia was evaluated bilaterally using the von Frey test as described previously (Chaplan, et al., 1994). To measure rat hind paw mechanical threshold, animals were placed into compartment enclosures in a test chamber with a framed metal mesh floor. Sensitivity to light touch was then determined using the up-down method of Dixon (1980) on days 7 and 14 post-surgery. Testing was initiated with a 2.0 g von Frey hair. In the absence of a paw withdrawal response, a stronger stimulus was used. In the event of paw withdrawal, a weaker stimulus was applied. Four additional stimulations with weaker/stronger hairs were performed following the initial response. The cut-off was set at the 15.0 g hair. The 50% g threshold was calculated using the method described in Chaplan et al. (1994) with a d value of 0.197.

*2.4.6 Chronic inflammatory pain model (complete Freund's adjuvant model)*

Under isoflurane anesthesia (5% in $O_2$, Abbott Laboratories, Montreal, QC, Canada), 100 µL of complete Freund's Adjuvant (CFA) emulsified 1:1 with saline 0.9% and containing 4 mg/mL of dessicated *Mycobacterium butyricum* was injected in the plantar surface of the right hind paw of male Sprague-Dawley rats. Twelve days after CFA administration, the rats received an i.t. administration of the vehicle (saline solution, 10% DMSO and 20% PEG) or the pepducin series (275 nmol/kg). On days 7 and 12 post-CFA, mechanical allodynia was assessed with the up-down method of Dixon (1980), as described above for the CCI animals.

***2.5 Detection of the fluorescent pepducin in rat dorsal root ganglion cells***

Male Sprague-Dawley rats were i.t. injected with the TAMRA-tagged pepducin (FP-001) at a 275 nmol/kg dose. Control animals were injected with vehicle alone (saline, 10% DMSO, 20% PEG4000). Twenty minutes post-injection, the rats were anesthetized with 5% isoflurane and euthanized. The dorsal root ganglia (DRGs) of the L4 to L6 lumbar vertebrae were rapidly isolated, post-fixed overnight in 4% paraformaldehyde at 4°C, and then cryoprotected in a 30% sucrose phosphate-buffered solution (PBS) at 4°C. Tissues were frozen at −20°C in Tissue-Tek O.C.T. Compound obtained by Sakura Finetek USA Inc. (Torrance, CA, USA) and then sectioned at 20 µm with a cryostat (Leica CM1860UV, Leica Biosystems, Concord, ON, Canada). Sections were washed twice in PBS for 10 min and incubated with DAPI staining (1:16000 dilution) for 10 min at room temperature. They were then rinsed twice with PBS for 10 min. The sections were mounted on SuperFost Plus slides (VWR, Mississauga, ON, Canada) and cover-slipped with Prolong Diamond Antifade Mountant (Thermofisher, Waltham, MA, USA). Images were acquired on a Leica DMI8 epifluorescence microscope (Leica Microsystems, Toronto, ON, Canada), using the same acquisition parameters (gain, exposure time). Captured images were analyzed using ImageJ software. Three animals per experimental condition were analyzed.

### *2.6 Data and statistical analysis*

All data obtained for this study was plotted onto graphs using Graphpad Prism 8 software (La Jolla, CA, USA) and represent the mean ± SEM of multiple independent experiments. In the BRET experiments, pepducins were tested a minimum of three times in duplicate. In the case of the cellular SPR assay, a minimum of six independent experiments were performed for each condition. For all *in vivo* behavioral tests, group sizes were determined as $n \geq 6$. Animals were randomly assigned to vehicle or to pepducin-treated groups by block randomization. Specific $n$ values are supplied in each figure legend. In order to compare the mean values between animal groups, two-way ANOVA tests were performed on behavioral data, with Tukey's or Sidak's correction for multiple comparisons. In the case where we compared the mean area under the curve between the vehicle and pepducin-treated groups, a Kruskal-Wallis test with Dunn's correction for multiple comparisons was performed. Statistical differences are depicted in the figures by asterisks (*, $p < 0.05$; **, $p < 0.01$; ***, $p < 0.001$; ****, $p < 0.0001$).

## 3. Results and Discussion

### *3.1 Pepducin design*

In order to determine whether or not pepducins targeting the NTS1 receptor may represent a valid strategy in pain relief, we designed and synthesized a pepducin series derived from the first intracellular loop domain (ICL1) of the NTS1 receptor. This series includes five main pepducins, a non-palmitoylated control, and a control pepducin with a scrambled peptide sequence. The peptide sequence for the first pepducin in our series (NTS1-ICL1-PP-001) corresponds to the full 14 amino acid sequence of the ICL1 domain of the human NTS1 receptor, as defined on the Uniprot database (ID: P30989, https://www.uniprot.org/uniprot/P30989). For the following pepducins in our series, this peptide sequence was sequentially truncated by increments of two C-terminal amino acids in order to identify a minimal biologically active fragment. The specific amino acid sequences for the full-length and truncated pepducins (**Supplementary Scheme S1**) are provided in **Table 1**. As there is no consensus regarding pepducin nomenclature, we identified these pepducins by a code detailing the target receptor (NTS1), the intracellular domain used as a template (ICL1), and the state of the peptide (i.e. PP = palmitoylated peptide, NP = non-palmitoylated peptide, FP = fluorophore-tagged palmitoylated peptide, **Supplementary Scheme S2**). Hereafter, the pepducins will be referred to in this article only by their abbreviated names (i.e. PP-001, PP-002, etc.).

The decision to use ICL1 as a starting point was primarily based on the fact that, contrarily to some of the other loops, it has a peptide sequence that is identical between the human, rat and mouse receptors, and we hypothesized that this would represent an experimental advantage when translating our findings between *in vitro* and *in vivo* models. Also, there are comparatively few

ICL1-based pepducins described in the literature to date (Zhang, et al., 2015a; O'Callaghan et al., 2012; Tressel, et al., 2011).

**Table 1: Peptide sequences of the ICL1-derived NTS1 pepducin series**

| Compound | Abbreviation | Amino Acid Sequence |
|---|---|---|
| NT(8-13) | -- | H- R R P Y I L -OH |
| NTS1-ICL1-PP-001 | PP-001 | Palmitoyl- A R K K S L Q S L Q S T V H -$NH_2$ |
| NTS1-ICL1-PP-002 | PP-002 | Palmitoyl- A R K K S L Q S L Q S T -$NH_2$ |
| NTS1-ICL1-PP-003 | PP-003 | Palmitoyl- A R K K S L Q S L Q -$NH_2$ |
| NTS1-ICL1-PP-004 | PP-004 | Palmitoyl- A R K K S L Q S -$NH_2$ |
| NTS1-ICL1-PP-005 | PP-005 | Palmitoyl- A R K K S L -$NH_2$ |
| NTS1-ICL1-NP-001 | NP-001 | H- A R K K S L Q S L Q S T V H -$NH_2$ |
| NTS1-ICL1-PP-SCR-001 | PP-SCR-001 | Palmitoyl- L V Q R L T A K S S K Q H S -$NH_2$ |
| NTS1-ICL1-FP-001 | FP-001 | TAMRA- Octadecanoyl- A R K K S L Q S L Q S T V H -$NH_2$ |

### *3.2 NTS1-derived pepducins activate NTS1-related G protein signaling pathways*

In order to determine whether these ICL1-derived pepducins were biologically active, we first explored their ability to engage the signaling pathways associated with NTS1 activation. Classically, the NTS1 receptor has been linked to the coupling and activation of the $G\alpha_q$ protein, known to initiate the cleavage of phosphatidyl inositol biphosphate (PIP2) via the activation of phospholipase C, leading to the intracellular release of inositol triphosphate ($IP_3$) and $Ca^{2+}$ (Amar, et al., 1987; Schaeffer, et al., 1995; Choi, et al., 1999). However, our team and others have shown that, when activated, NTS1 may engage multiple signaling pathways, including the $G\alpha_s$, $G\alpha_i$, $G\alpha_o$, and $G\alpha_{13}$ protein signaling pathways, as well as the recruitment of β-arrestins 1 and 2 (Besserer-Offroy, et al., 2017; Hwang, et al, 2010; Gailly, et al., 2000; Oury-Donat, et al., 1995). In this study, we thus screened our pepducin series in $G\alpha_q$ and $G\alpha_{13}$ signaling pathways, specifically. G protein activation was monitored using biosensors based on Bioluminescence Resonance Energy

Transfer (BRET) technology, which consist of Gα subunits tagged with a *Renilla* luciferase (RlucII) donor and Gγ subunits tagged with a GFP10 acceptor.

Here, the experimental paradigm was to stimulate CHO-K1 cells transiently expressing the BRET-based biosensors and the hNTS1 receptor with either NT(8-13) (1 μM) or the ICL1-derived pepducins (10 μM), and to monitor G protein dissociation over a 30-minute time period. Note that the C-terminal hexapeptide fragment of NT, NT(8-13), corresponds to the minimal biologically active fragment (St-Pierre, et al., 1981), and has been reported to promote NTS1 signaling in an identical fashion to the full-length NT (Besserer-Offroy, et al., 2017). In **Figure 1A** and **1B**, we show G protein activation for each condition 15 minutes post-stimulation. We observed partial $G\alpha_q$ protein activation by the pepducins PP-001 to PP-005, ranging in magnitude from 30 ± 8% activation (PP-001) to 52 ± 15% activation (PP-005). We observed a greater $G\alpha_{13}$ protein activation, ranging in values from 33 ± 2% (PP-001) to 98 ± 3% activation (PP-003). Importantly, treatment with NP-001, PP-SCR-001 or palmitate alone resulted in no G protein activation for both assays. Furthermore, no G protein activation was observed in mock CHO-K1 cells lacking hNTS1 receptor expression (**Supplementary Figure S1**).

These findings suggest that our NTS1-derived pepducins can act as allosteric agonists of the receptor by promoting NTS1-related signaling. Admittedly, this agonist effect is only partial, compared to NT(8-13), when tested at a high concentration (10 μM). We chose to screen our pepducins at this concentration, as the reported potencies for pepducin agonists are generally much lower than for orthosteric agonists. For example, the CXCR4-derived pepducin, ATI-2341, has been shown to stimulate the $G\alpha_{i1}$ signaling pathway with an $EC_{50}$ of 208 ± 70 nM, while CXCR4's endogenous ligand (SDF-1) promotes $G\alpha_{i1}$ signaling with an $EC_{50}$ of 0.25 ± 0.06 nM (Quoyer, et al., 2013). Likewise, the β2AR-derived ICL3-9 pepducin produces cAMP with a 4.7 ± 0.1 μM potency (Carr III et al., 2014), whereas β2AR's endogenous ligand (isoproterenol) promotes cAMP production with an $EC_{50}$ of 8.23 ± 0.15 nM (van der Westhuizen, et al., 2014). To our knowledge, the most potent pepducin described so far is ICL1-9, also derived from β2AR, which recruits β-arrestins with an $EC_{50}$ of 96 ± 14 nM (albeit only with partial efficacy) (Carr III, et al., 2016), compared to 20 ± 5 nM for isoproterenol (Hamden, et al., 2005). Other pepducins from the same ICL1-based β2AR pepducin series presented with $EC_{50}$ values of 1-2 μM (ICL1-4 = 1.9 ± 0.5 μM; ICL1-11 = 1.7 ± 0.5 μM and ICL1-20 = 1.1 ± 0.3 μM). When interpreting these functional potencies, the particular mode of action of pepducins should be taken into account, as pepducins must first cross the membrane barrier in order to modulate receptor activity. This mechanism of action by which pepducins passively flip between the outer and inner leaflets of the cell membrane in order to enter into the cell has been confirmed by numerous studies. In 2007, Wielders and colleagues used a FRET-based assay and differentially labeled phospholipids to demonstrate that PAR1 pepducins are present in the outer and inner leaflets of the plasma membrane (Wielders, *et al.,* 2007). Tsuji and colleagues developed their own FRET probes based on the PAR1-derived P1pal-13 pepducin, with a fluorescent switch component that enabled them to observe transbilayer pepducin movements using live cell imaging (Tsuji, *et al.,* 2013). As this process is passive and bidirectional, we cannot know what proportion of the dose administered functionally interacts with the target receptor at any given time-point. Determinations of pepducins' potency and efficacy, therefore, cannot depend solely on the affinity of a given pepducin for its cognate receptor or on factors such as cell-type and environment, but must also include considerations such as the rate at which pepducins cross the membrane barrier, and how they are distributed between the inner and outer leaflets of the cell membrane.

### *3.3 NTS1-derived pepducins induce NTS1-dependent dynamic mass redistribution responses in a cell monolayer*

While the BRET experiments indicated that our pepducins could activate NTS1-related signaling pathways, we recognized that NTS1 has a large signaling repertoire, and that establishing pepducin signaling profiles by monitoring the activation of individual signaling pathways would be a complex task. We therefore sought to establish global pepducin response profiles by measuring a cellular response that results from the integration of all molecular events within the cell. Consequently, we monitored the whole-cell integrated responses of a CHO-hNTS1 cell monolayer by surface plasmon resonance (SPR), in response to 10 μM of NT(8-13) and of each of the pepducins in our series (**Figure 2**). Briefly, this optical sensing technique detects changes in the refractive index of a laser as it passes through a prism and is reflected off a gold-plated substrate, upon which living cells have been seeded (Chabot, et al., 2009). Molecular reorganization and corresponding dynamic mass redistribution (DMR) that occur in the basal portion of the cell monolayer are detected as variations in reflectance (RVU), which correspond to cellular activity such as cytoskeleton remodeling, cell spreading, and cell-cell or cell-substrate adhesion or tension (Cuerrier, et al., 2008). This non-invasive and label-free technique therefore offers the advantage of integrating all the molecular events associated with the activation or inhibition of cell signaling activity.

As can be observed in **Figure 2A**, the pepducin-treated conditions presented with SPR traces that differ greatly from NT(8-13)'s, and from each other's. Whereas the SPR trace for NT(8-13), previously reported by our team (Besserer-Offroy, et al., 2017), is characterized by an initial decline in reflectance (-5 ± 1 RVU) and a subsequent increase which, after a brief plateau, reaches a maximum reflectance of 27 ± 5 RVU, the SPR traces monitored for our pepducin series are quite different. In the case of PP-001, we observed a robust increase in reflectance reaching a maximal amplitude of 80 ± 13 RVU at 400 seconds post-stimulation, followed by a gradual decline in reflectance until the end of the experiment. In the case of PP-002 and PP-003, we observed a more gradual increase in reflectance which reached values of 90 ± 10 RVU and 126 ± 20 RVU, respectively. In the case of PP-004 and PP-005, presented in **Figure 2B**, a high initial rate of change in reflectance was observed, reaching high amplitude plateaus (309 ± 12 RVU and 241 ± 12 RVU, respectively) within 400 seconds that lasted until the end of the assessment. Based on these traces, the pepducin series can be divided into three distinct profile groups (PP-001; PP-002 and PP-003; PP-004 and PP-005). Due to the integrative nature of the assay, it is difficult to delineate the contribution of particular cellular or signaling events associated to the observed reflectance profiles. However, the results indicate that there is greater global signaling activity for PP-004 and PP-005, compared to PP-002 and PP-003. Also, these results could point toward a more specific signaling activity for PP-001. Nevertheless, this SPR experiment confirmed that the pepducins are biologically active and induce cellular responses in a NTS1-expressing cell population that differ from those of NT(8-13), suggesting that they may stabilize alternate receptor conformations than the native ligand, potentially producing very different outcomes.

As pepducins are lipidated peptides, they could induce cellular toxicity, we thus assessed cell viability after a prolonged exposure to PP-001. As shown in **Supplementary Figure S2**, after 16 hours of incubation with 10 μM of PP-001, cell viability was comparable to the vehicle treated cells suggesting that, despite their physicochemical properties, pepducins do not reduce cell viability.

Importantly, control experiments were performed in which CHO-hNTS1 cells received 10 μM of NP-001 and PP-SCR-001, and in which mock CHO-K1 cells received 10 μM of NT(8-13) or the PP-001 pepducin (**Figure 2C-F**). The SPR signal for each of these conditions remained steady, and no detectable changes in reflectance were observed. This key result, which demonstrates that receptor expression is required for pepducin action, correlates with what is known for the great majority of pepducins. Indeed, pepducins are reported to be receptor-dependent; both PAR4- and CXCR4-targeting pepducins, for example, lose their activity when the target receptor is absent (Covic, et al., 2002; Tchernychev, et al., 2010). Thus, while much about the pepducin mode of action remains unclear, they do appear to act via a direct pepducin-receptor interaction. This conclusion is notably supported by work from Janz and colleagues, who used a photochemical cross-linking approach to confirm that the ATI-2341 pepducin is a direct binding partner of its cognate receptor, CXCR4 (Janz, et al., 2011). Other recent studies have instead successfully attempted to visualize pepducin-induced changes in receptor conformation using purified receptor systems conjugated to environmentally sensitive fluorophores, such as monobromobimane (Carr III, et al., 2014; Zhang, et al., 2015b). Thus, the authors were able to determine that the β2AR-derived pepducin ICL3-9 induces a unique conformational change in the TM6 region of the receptor, one that differs from that of the endogenous agonist isoproterenol (Carr III, et al., 2014). Conversely, the PAR1-derived pepducin P1pal-19 appears to induce a similar conformational change to PAR1's TM5 as that induced by the endogenous agonist thrombin (Zhang, et al., 2015b). These indications of receptor conformational changes reinforce the idea that the target receptor is critical to the pepducin mode of action. We are aware of only one pepducin, a β2-adrenergic receptor-derived pepducin named ICL3-8, that is reported to promote $G\alpha_s$-downstream signaling independently of the cognate receptor, and its mechanism of action is yet undetermined (Carr III, et al., 2014).

### *3.4 PP-001 and PP-005 significantly reduce acute pain*

Having demonstrated that our pepducins were able to engage G-protein-mediated signaling pathways and to produce distinct DMR responses associated with NTS1 activation, we next assessed the antinociceptive potential of our pepducins in the tail-flick acute pain assay. We measured the latency in the tail withdrawal response of healthy adult male rats to a noxious thermal stimulus (a 52ºC water bath) following i.t. injection of the pepducins. This pain model is generally used as a tool to screen for analgesics, of which the most promising are then brought along into more clinically relevant pain models. Thus, we used this assay to evaluate the analgesic effects of our pepducin series by testing PP-001, PP-003 and PP-005 (each presenting a distinct SPR profile) at a dose of 275 nmol/kg (**Figure 3A**). Please note that the pepducins were not degraded for at least 60-min in rat cerebrospinal fluid, as determined in an *in vitro* metabolic stability assay (**Supplementary Figure S3**). As can be seen, administration of the full-length pepducin (PP-001) resulted in a significant increase in the tail withdrawal latency, compared to vehicle-treated animals, that began immediately following i.t. injection and lasted for 60 min. In contrast, PP-003 was ineffective at increasing the reaction time of the rats in the tail-flick test. I.t. delivery of PP-005 produced a significant antinociceptive effect that was weaker than PP-001's. Notably, the antinociceptive actions of PP-001 and PP-005 were not accompanied by any observable side effects. Injections of NP-001 or PP-SCR-001 at the same dose did not result in a significant increase in the tail withdrawal latency (**Figure 3B**). In **Figure 3C**, we provide the response amplitude values at the 40 min-timepoint post-injection, where we can observe significant effects for PP-001 (33 ± 6% MPE) and PP-005 treatment (17 ± 4% MPE), compared to the vehicle condition (4 ± 2% MPE). NP-001 and PP-SCR-001 had no significant effects (10 ± 3% and 6 ± 2% MPE, respectively).

Again, we should highlight the fact that the antinociceptive effects observed here occur when the pepducins are administered at 275 nmol/kg. For PP-001, this corresponds to 500 μg/kg. This high dose was chosen to reflect the shift in potency we expect for our pepducins, relative to traditional orthosteric ligands. Indeed, we previously reported the design of a NT(8-13) analogue, JMV2007 (compound 6), that binds NTS1 with an $IC_{50}$ of 0.02 ± 0.002 nM, and increased tail-withdrawal latencies when administered i.t. at a 30 μg/kg dose (Fanelli, et al., 2015). Likewise, we recently reported a novel series of macrocyclic NT(8-13) analogues and therein characterized a compound with a 15 ± 2 nM ($K_i$) binding affinity for NTS1, which we tested at various doses in our acute pain model. We found that it presented with an $ED_{50}$ of 4.63 μg/kg and reached maximal efficacy at 10 μg/kg (Sousbie, et al., 2018). In comparison, the dose at which PP-001 was administered represents a 15 to 50-fold increase (relative to 10 and 30 μg/kg, respectively). As indicated previously, pepducins typically present with micromolar potencies in cell-based assays, whereas orthosteric ligands may often present with nanomolar potencies. Thus, we might reasonably expect a 100-fold shift in potency, simply as a result of the nature and mode of action of pepducins. Here, the fact that we clearly observe pepducin-induced antinociception in this model, at this dose, suggests that PP-001 and PP-005 are very effective at relieving pain. Accordingly, these pepducins were retained for further study into their analgesic properties.

### *3.5 PP-001 attenuates formalin-induced nociceptive behaviors in the tonic pain model*

We next assessed the antinociceptive potential of these pepducins by using the formalin test as a model of tonic (or persistent) pain. As expected, intraplantar injection of formalin into the right hind paw of saline-pretreated rats induced a biphasic time-dependent increase in pain scores (**Figure 4A**). However, rats pre-injected with PP-001 (i.t., 275 nmol/kg) exhibited significantly lower pain scores during the inflammatory phase. This reduction in spontaneous nociceptive-related behaviors is also reflected in the total area under the curve (AUC) values calculated for the prolonged inflammatory phase (21-60 min) (**Figure 4B**), where we observe a 63 ± 14% decrease in PP-001-treated animals compared to the vehicle. PP-005 had a lesser, though still significant, antinociceptive effect in the tail-flick test; however, it proved to be ineffective at relieving pain in the formalin model (**Figure 4C-D**).

In an effort to ensure that the antinociception observed by PP-001 is mediated *in vivo* by the target receptor NTS1 and not the result of non-specific actions on other membrane proteins, PP-001 was co-administered with the NTS1 antagonist SR48692 at 10 µg/kg dose (i.e. 17 nmol/kg). Here, the co-administration of this NTS1 antagonist entirely reversed the PP-001-induced antinociception (**Figure 4B**). While this experiment is not sufficient to conclude on the specificity of this pepducin for its target receptor, it strongly suggests that the antinociceptive effects of PP-001 are indeed mediated via NTS1. This is further supported by our earlier *in vitro* findings showing that PP-001's activity in the SPR experiment is NTS1-dependent. This is an important aspect to consider, as there is some debate in the literature regarding pepducin target specificity and selectivity. One of the advantages initially attributed to the pepducin approach was the potential for high selectivity, as multiple pepducins were reported to distinguish between subtypes of a same receptor family (Covic, et al., 2002; Janz, et al., 2011). In certain cases, however, pepducins have been found to target multiple 7TMRs, such as the PAR4-derived pepducin P4pal-10 that inhibits agonist-induced responses for a range of Gαq-coupled receptors, not only PAR4 (Carr III, et al., 2016b).

### *3.6 PP-001 reverses mechanical hypersensitivity in rat models of chronic neuropathic and inflammatory pain*

Subsequently, the analgesic potential of the PP-001 pepducin was assessed in rat models of chronic neuropathic and inflammatory pain. The *Chronic Constriction Injury* (CCI) model, which consists in the loose ligation of the rat sciatic nerve, was first used to mimic chronic neuropathic pain. The development of mechanical allodynia was assessed using the von Frey test before as well as 7 and 14 days post-surgery. As shown in **Figure 5A**, the CCI rats developed tactile allodynia by day 7 post-surgery, as the mechanical thresholds decreased from a baseline of 15 g to 5.1 ± 0.4 g, corresponding to 66 ± 3% paw withdrawal threshold reduction, compared to preoperative values. On day 14 post-surgery, i.t. injection of PP-001 produced robust anti-allodynic effects, reaching up to 84 ± 16% of pain relief at maximum effect.

We further evaluated the analgesic efficacy of PP-001 in a model of chronic inflammatory pain following *Complete Freund's Adjuvant* (CFA) intra-plantar injection into the right hind paw. The heat-inactivated mycobacteria (*M. butyricum)* present in the CFA provoke a strong immune response, which subsequently results in hypersensitive reactions to mechanical stimuli. On day 7 post-CFA injection, the paw withdrawal thresholds in response to tactile stimuli greatly decreased from 15 g in naïve animals to 3.0 ± 0.2 g in CFA rats (i.e. 80 ± 2% reduction in paw withdrawal thresholds). Importantly, i.t. injection of PP-001 (275 nmol/kg) significantly reversed the mechanical inflammatory pain hypersensitivity on day 12 post-CFA injection (60 ± 14%), thus confirming the analgesic activity of PP-001 in a variety of pain paradigms (**Figure 5B**).

Prior to this work, NTS1 was untargeted by pepducin technology, and pain research was an area largely untouched. Indeed, very few studies have been published in which pepducins were administered in a pain paradigm. In 2009, McDougall and colleagues sought to decipher the role of PAR4 in joint pain and inflammation and found that the PAR4-derived pepducin p4pal-10 could inhibit the thermal and mechanical hyperalgesia produced by intra-articular injection of a PAR4 agonist (McDougall, et al., 2009, Russell, et al., 2009). Similarly, Annahazi and colleagues used p4pal-10 to investigate the role of PAR4 on visceral pain produced by colorectal infusion of fecal supernatants originating from patients with ulcerative colitis (UC) or irritable bowel syndrome (IBS) (Annahazi, et al., 2009). In this experimental condition, they found that p4pal-10 treatment exacerbated visceral sensitivity in UC-infused rats. We are aware of only one other study to use a pepducin in a pain context. In this study, a pepducin derived from the ICL2 of the delta opioid receptor (DOP), C11-DOPri2, reversed the analgesic action of deltorphin II in CFA-treated rats (Beaudry, et al., 2015). In each of these studies, the pepducins were used to inhibit agonist-mediated effects, and thus have been described as antagonists of their cognate receptors. However, due to the allosteric nature of pepducins, we deem it more appropriate to describe them as negative allosteric modulators. Regardless, the pepducins were used as pharmacological tools to decipher the role of their target receptors in a specific pain context, or to better understand receptor-effector interactions. While pepducin treatment was not the focus of these studies, they nevertheless provided a basis for considering that pepducins might effectively modulate the pain response and thus might represent a valid strategy in pain-relief. Our findings lend further credence to this idea, as we report a pepducin (PP-001) that effectively reduces nociceptive behaviors in multiple pain models. Notably, this pepducin-mediated effect is maintained even in chronic pain states that are very challenging to treat.

***3.7 FP-001 reaches the dorsal root ganglia after i.t. delivery***

It has previously been reported that the NTS1 receptor is expressed in the laminae I and II of the spinal cord dorsal horn, as well as in subpopulations of small- and medium-sized neurons in the dorsal root ganglia (DRGs) of naïve rats (Roussy, et al., 2008). As the DRGs house the cell bodies of primary afferents critical to pain sensation and transmission, we questioned whether or not PP-001 reached the primary sensory neurons following lumbar i.t. injection. In order to investigate this, we designed an ICL1-derived pepducin tagged with the TAMRA fluorophore in the N-terminal position, which we injected (i.t.) into rats at 275 nmol/kg. Twenty minutes post-injection, the lumbar DRGs were isolated and fixed, and we observed TAMRA-associated fluorescence via confocal microscopy. As can be seen in **Figure 6**, a diffuse TAMRA staining was detected in the different subtypes of sensory neurons (small, medium and large) at the level of the L5-L6 vertebrae of pepducin-treated animals. This fluorescence was absent in vehicle-treated animals (**Supplementary Figure S4**). It is worth noting that all subtypes of DRG neurons are labeled after FP-001 injection, this can be explained by the fact that a pepducin anchors itself in any cell membrane regardless of the presence of the receptor but only signals when its target receptor is present. Furthermore, no staining was observed in the superficial laminae of the dorsal horn of the spinal cord. In order to ensure that the labeling of PP-001 with the TAMRA fluorophore did not alter its antinociceptive effects, we tested FP-001 in our acute pain model, and measured tail withdrawal latencies to a noxious thermal stimulus. We observed a similar increase in the tail withdrawal latency in FP-001-treated rats as with PP-001, with a maximal possible effect of 40 ± 9% at 40 min post-injection. This data may suggest that when injected i.t., PP-001 modulates the pain response directly at the primary afferent nociceptors.

In this study, we applied pepducin technology to the study and treatment of chronic pain, using NTS1 as a therapeutic target. We recognized that pepducins, by their very particular mode of action, offer unique opportunities in drug discovery. Indeed, drug development in recent years has been profoundly impacted by concepts such as biased agonism and allostery (Kenakin, 2012), suggesting that we might be able to fine-tune 7TMR responses in order to produce desired therapeutic outcomes and limit undesirable effects (Kenakin, A Pharmacology Primer, 2009). Pepducins that act allosterically at the intracellular surface of 7TMRs are particularly interesting from this viewpoint, as they may access distinct receptor conformations unavailable to traditional orthosteric, or even extracellular allosteric ligands (Carr III, et al., 2014).

The NTS1-derived pepducins presented here appear to act as allosteric agonists of NTS1, as evidenced by the BRET experiments monitoring G protein activation. However, the very distinct pepducin response profiles observed in the SPR assay indicate that their global effect on a cell population is very different to that produced by orthosteric activation of NTS1. Presumably, they may stabilize NTS1 receptor conformations that lead to distinct physiological responses, for which only a subset of the effects associated with NTS1 activation are produced (analgesia, hypotension, hypothermia, ileum contraction and relaxation, etc.). This hypothesis is supported by the observation that, contrarily to other NTS1 agonists, i.t. administration of PP-001 does not appear to produce hypothermia in male Sprague-Dawley rats. Indeed, spinal delivery of the NTS1 agonist PD149163 (16.5 nmol/kg) leads to a $1.6 \pm 0.2^{o}$C drop in mean body temperature, whereas PP-001 at 275 nmol/kg does not induce hypothermia (**Supplementary Figure S5**). Potentially, therefore, these pepducins could not only be of great therapeutic interest, but valuable as tools to decipher the signaling pathways responsible for NTS1's different physiological effects, the mechanisms behind many of which remain unclear.

However, it is evident that the *in vitro* and *in vivo* experiments we performed in this study represent a very preliminary characterization of the pepducins' actions. To truly understand how these pepducins modulate the NTS1 receptor, additional experiments should be performed to further explore both their cellular and physiological effects. As mentioned previously, it is also feasible that the pepducins in our series do not solely target NTS1, but also act on other 7TMRs and/or cellular effectors. However, the fact that PP-001 does not exhibit any activity when the receptor is absent in the SPR experiment, as well as the observation that co-administration of PP-001 with the NTS1 antagonist SR48692 reverses its antinociceptive response, suggests that these actions are indeed mediated via its target receptor NTS1.

## 4. Conclusions

In this proof-of-concept study, we sought to determine whether pepducin technology is applicable to the study and treatment of chronic pain. We chose the NTS1 receptor as a novel template from which to design a pepducin series. Our findings indicate that these pepducins are biologically active, can engage NTS1-related signaling pathways at concentrations of 10 µM and can induce distinct DMR response profiles in a NTS1-expressing cell population. Most importantly, we found that the PP-001 pepducin can effectively relieve pain in acute, tonic and chronic pain models when injected (i.t.) at a 275 nmol/kg dose. This study represents a first foray into the field of pepducin-mediated pain-relief and may help inform the development of novel analgesics that could better cater to the needs of chronic pain patients.

**Acknowledgements**

RLB and MC are supported by research scholarships awarded by the Fonds de recherche du Québec – Santé (FRQ-S) and by the Faculty of Medicine and Health Sciences (FMSS) of the Université de Sherbrooke. ÉBO is supported by FRQ-S and CIHR research fellowships. PS holds a Canada Research Chair in Neurophysiopharmacology of Chronic Pain. Drs. M. Bouvier, T. Hebert, S.A. Laporte, G. Pineyro, J.-C. Tardif, E. Thorin and R. Leduc (CQDM Team) are also acknowledged for providing us with the G protein BRET-based biosensors.

**Funding sources**

This work was supported by a Fonds de recherche du Québec – Nature et technologies (FRQ-NT) team research grant (2018-PR-207951), a Canadian Institutes of Health Research (CIHR) grant (FDN-148413), and a Centre d'excellence en neurosciences de l'Université de Sherbrooke (CNS) grant (Pilot Project Grant Scheme).

**Conflicts of interest**

The authors declare no competing financial interests.

**Author contributions**

Conception and design of study: ÉBO, CEM, and PS.

Supervision and direction of the study: ÉBO, CEM, JML, MG, and PS.

Acquisition of data: RLB, MC, SL, KB, and PS.

Analysis and interpretation of data: RLB, ÉBO, CEM, JML, and PS.

Contributed new reagent and analysis tools: CEM, MS, and ÉM.

Responsible for securing funds: ÉBO, CEM, ÉM, MG, and PS

Writing first draft of the manuscript: RLB.

Revision of the manuscript before submission: ÉBO, JML, ÉM, MG, and PS.

All the authors approved the version of the manuscript to be published.

## Figures & Legends

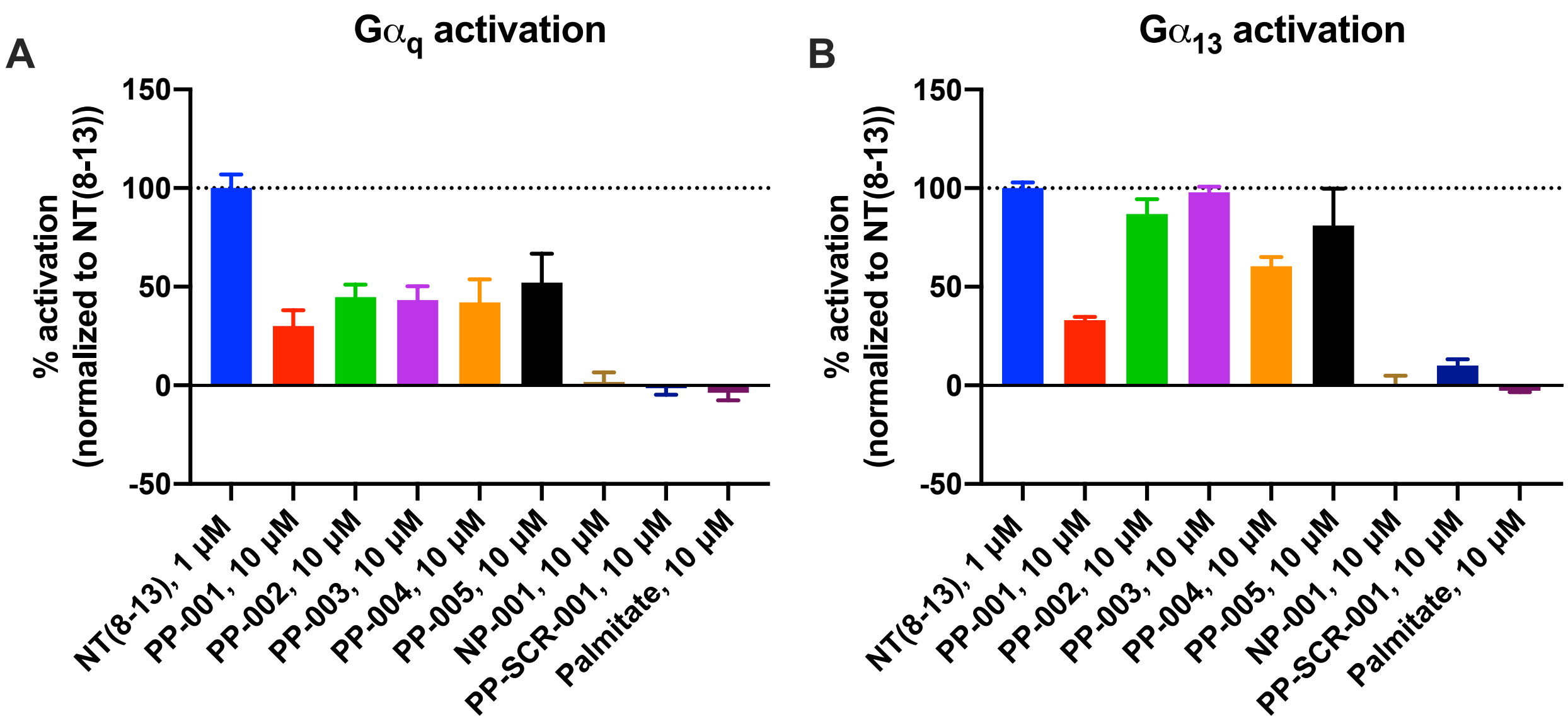

**Figure 1. Activation of G protein signaling pathways in response to pepducin treatment, monitored by BRET[2].** $G\alpha_q$ (**A**) and $G\alpha_{13}$ (**B**) protein activation in CHO cells transiently transfected with hNTS1 and the BRET[2]-based biosensors, 15 min post-stimulation with NT(8-13) (1 µM), the ICL1-derived hNTS1 pepducins or palmitate (10 µM). BRET[2] ratios were normalized according to NT(8-13); values for non-treated cells were set as 0% pathway activation, and those for cells stimulated with 1 µM NT(8-13) were set as 100 % pathway activation. Data represent mean ± SEM, *n* = 3, performed in duplicate.

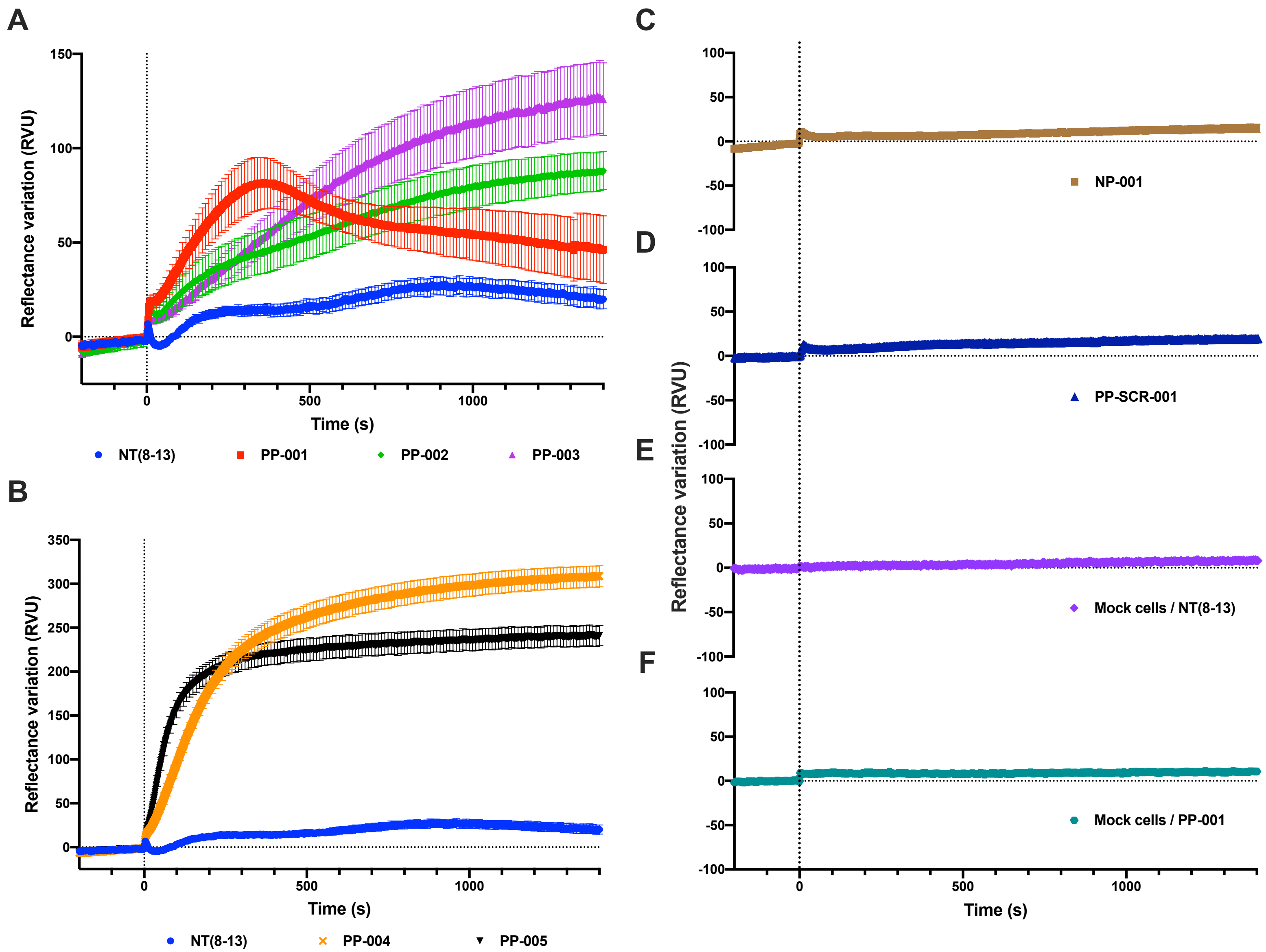
A
Reflectance variation (RVU)
150
100
50
0
0
500
1000
Time (s)
NT(8-13)
PP-001
PP-002
PP-003
B
Reflectance variation (RVU)
350
300
250
200
150
100
50
0
0
500
1000
Time (s)
NT(8-13)
PP-004
PP-005
C
100
50
0
-50
-100
NP-001
D
100
50
0
-50
-100
PP-SCR-001
E
Reflectance variation (RVU)
100
50
0
-50
-100
Mock cells / NT(8-13)
F
100
50
0
-50
-100
Mock cells / PP-001
0
500
1000
Time (s)

**Figure 2. Whole-cell integrated responses to pepducin treatment, monitored by surface plasmon resonance.** Variations in reflectance (normalized response) observed by surface plasmon resonance in CHO-K1 cells stably expressing hNTS1, treated with NT(8-13) and each of the ICL1-derived hNTS1 pepducins (**A-B**), or with the controls NP-001 (**B**) and PP-SCR-001 (**C**) at a 10 μM concentration. Variations in reflectance observed in mock CHO-K1 cells (absent hNTS1 receptor) treated with NT(8-13) (**D**) or with PP-001 (**E**) at 10 μM. SPR was monitored for a 30-min period, during which time a steady signal was acquired for at least 10 minutes prior to compound stimulation. Data represent mean ± SEM, *n* = 6. Dotted line at x=0 represents time of compound stimulation.

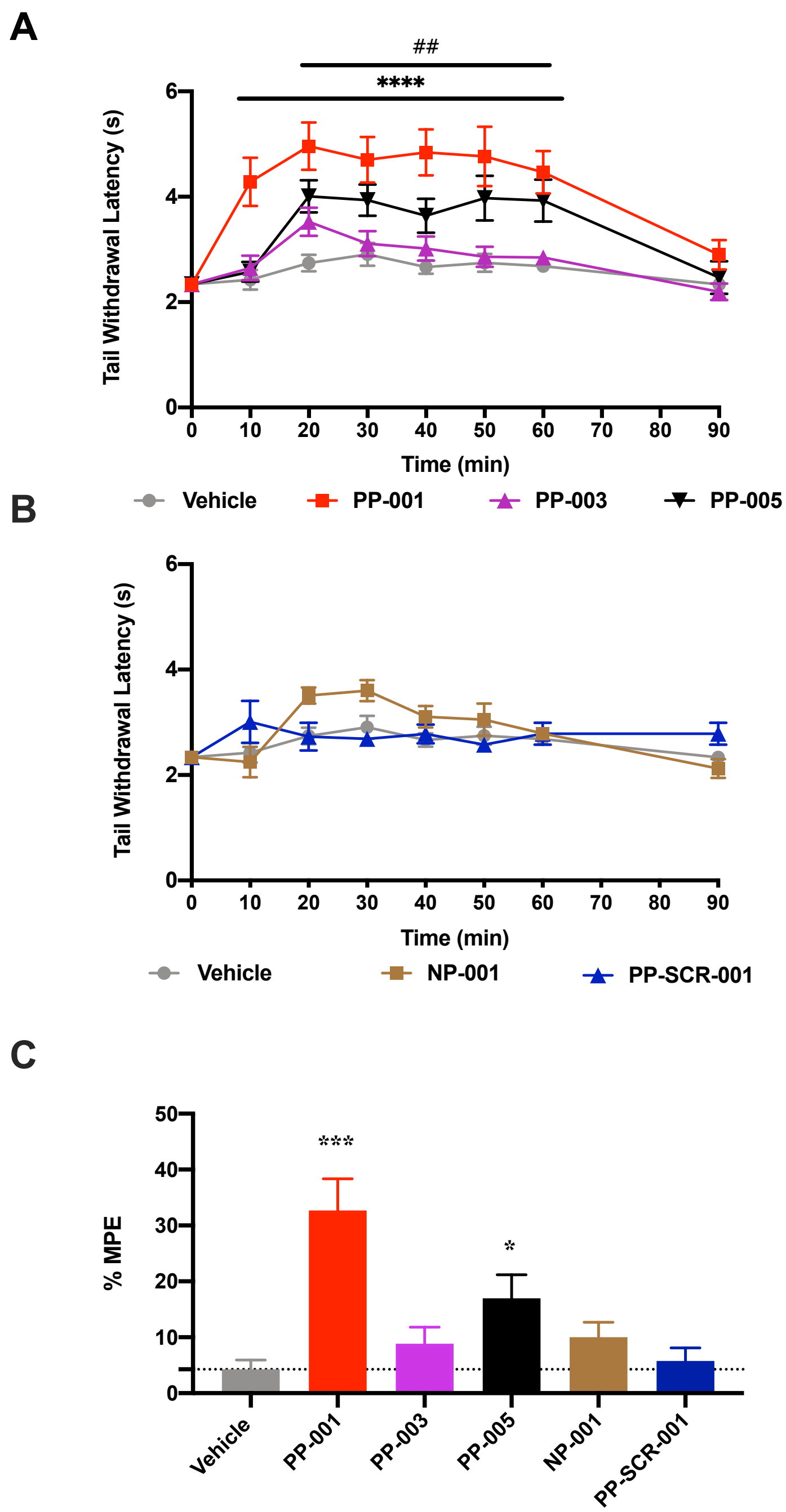

**Figure 3. Effects of ICL1-derived pepducins on rat acute pain (tail-flick test).** Tail withdrawal latencies observed over a 90-min period in male Sprague-Dawley rats injected (i.t.) with pepducins PP-001, PP-003, or PP-005 (**A**), with control pepducins NP-001 and PP-SCR-001 (**B**), or with vehicle (saline, 10% DMSO, 20% PEG4000). All pepducins and controls were injected at 275

nmol/kg. The percentage of *maximal possible effects* (% MPE) of tail withdrawal were calculated at 40 min post-injection, for each condition (**C**). Data represent mean ± SEM, $n = 6$-12 rats per group. A two-way ANOVA with Tukey's correction for multiple comparisons was performed (**A-B**). A Kruskal Wallis test with Dunn's correction for multiple comparisons was also performed (**C**). Asterisks denote statistical differences between the vehicle and PP-001-treated group: *, $p < 0.05$; **, $p < 0.01$; ***, $p < 0.001$. Sharps denote statistical differences between the vehicle and PP-005-treated group: ##, $p < 0.01$.

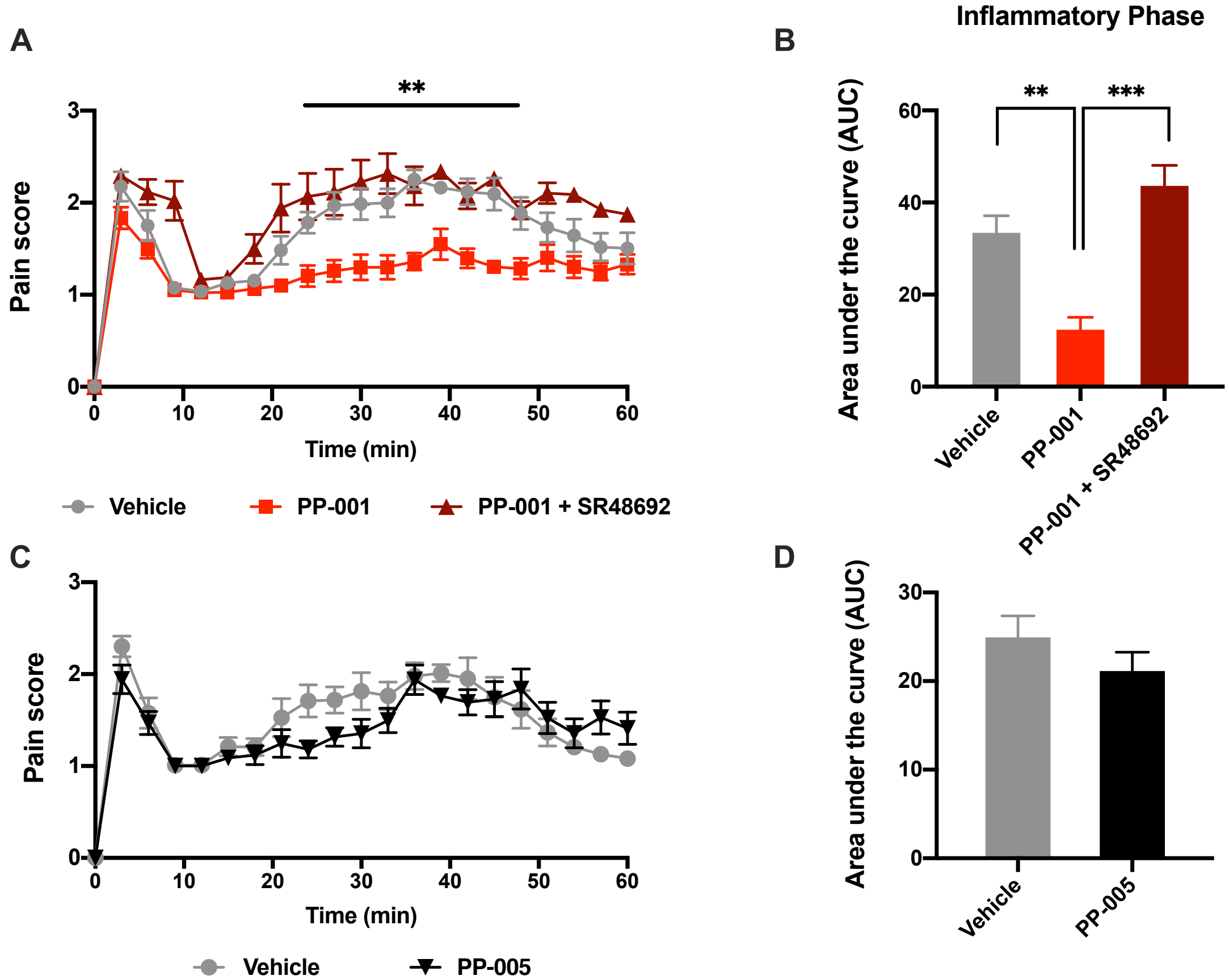

**Figure 4. Analgesic effects of ICL1-derived pepducins PP-001 and PP-005 on rat tonic pain (formalin test)**. Nociceptive responses, expressed as weighted pain scores, of male Sprague-Dawley rats following subcutaneous injection of formalin in the right hind paw. Five minutes prior to formalin administration, rats were injected (i.t.) with vehicle (saline, 10% DMSO, 20% PEG4000), with PP-001 (275 nmol/kg), with PP-001 (275 nmol/kg) and NTS1 antagonist SR48692 (10 µg/kg) (**A**), or with PP-005 (275 nmol/kg) (**C**). The cumulative nociceptive response, expressed as mean of area under the curve (AUC), observed during the second (inflammatory) phase of the formalin test (21-60 min) (**B, D**). Data represent mean ± SEM, $n$ = 6-12 rats per group. A two-way ANOVA with Tukey's correction for multiple comparisons (**A, C**), and a Kruskal-Wallis test with Dunn's correction (**B, D**), were performed. **, $p < 0.01$; ***, $p < 0.001$.

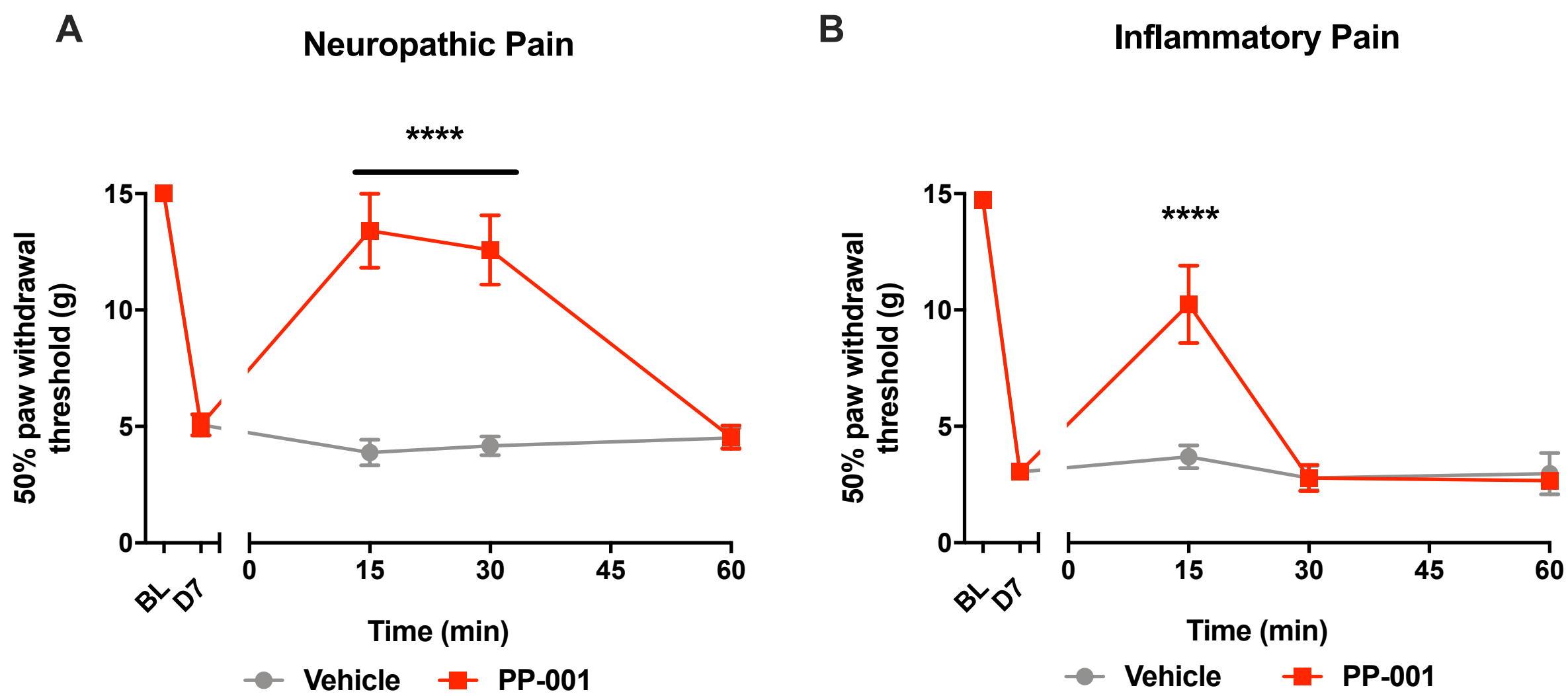

**Figure 5. Analgesic effects of ICL1-derived pepducin PP-001 on rat chronic neuropathic and inflammatory pain.** Mechanical allodynia thresholds in male Sprague-Dawley rats following ligation of the sciatic nerve (*Chronic Constriction Injury (CCI) model*) (**A**) or following hind paw injection with Complete Freund's Adjuvant (*Complete Freund's Adjuvant (CFA) model*) (**B**), as determined by von Frey test. Testing was performed prior to intervention (BL), and on days 7, 12 (CFA) or 14 (CCI) post-intervention. Rats were injected (i.t.) with PP-001 (275 nmol/kg) or with vehicle (saline, 10% DMSO, 20% PEG4000) on Days 12/14 and observed over a 1-hour period. Thresholds depicted correspond to the operated (ipsilateral) limb. Data represent mean ± SEM, with a $n$ = 5 (CCI) or $n$ = 6 (CFA) rats per group. A two-way ANOVA with Sidak's correction for multiple comparisons was performed. ****, $p < 0.0001$.

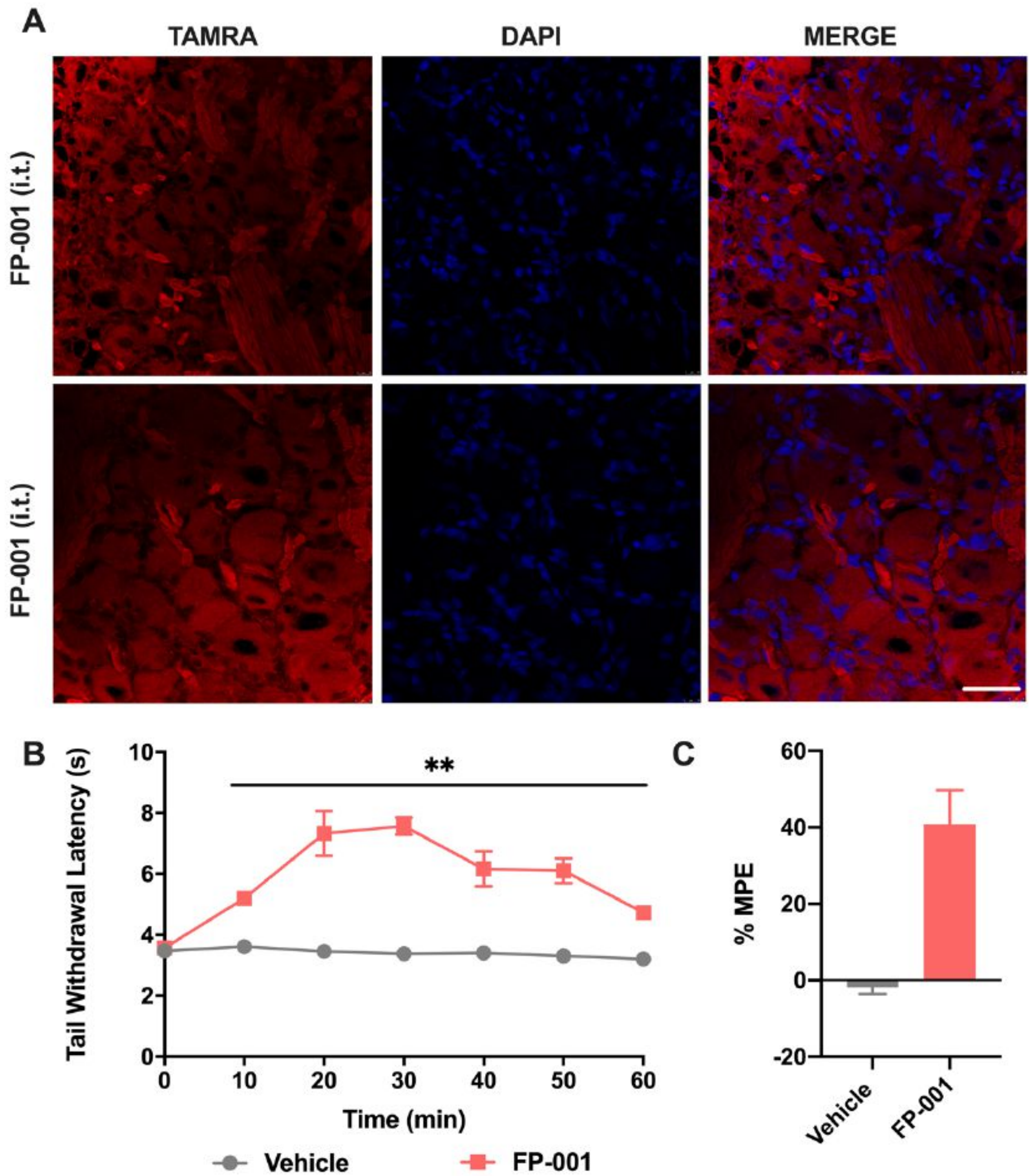

**Figure 6. Detection of fluorescent pepducin FP-001 in rat dorsal root ganglion neurons following i.t. injection.** Confocal microscopy images of DRG neurons extracted at the level of the L6 vertebrae from rats injected (i.t.) with the TAMRA-labeled pepducin FP-001 (275 nmol/kg, *n* = 5 rats). TAMRA-associated fluorescence was visualized in the red filter channel. DAPI (blue) was used as a nuclear counterstain. Confocal images were acquired under 20x magnification

(upper panel; Scale bar = 80 µm) and 40x magnification (lower panel; Scale bars = 40 µm). (**A**). Tail withdrawal latencies observed over a 60-min period in male Sprague-Dawley rats injected (i.t.) with FP-001 (275 nmol/kg) or with vehicle (saline, 10% DMSO, 20% PEG4000) ($n$ = 3 rats) (**B**). The percentage of *Maximal possible effect* (% MPE) of tail withdrawal were calculated at 40 min post-injection, for each condition (**C**). Data represent mean ± SEM. A two-way ANOVA with Sidak's correction for multiple comparisons was performed (**B**). **, $p < 0.01$.

# Cell-penetrating pepducins targeting the neurotensin receptor type 1 relieve pain

## Supplementary Information

Rebecca L. Brouillette[1‡], Élie Besserer-Offroy[2‡*], Christine E. Mona[3], Magali Chartier[1], Sandrine Lavenus[1], Marc Sousbie[1], Karine Belleville[1], Jean-Michel Longpré[1], Éric Marsault[1], Michel Grandbois[1], Philippe Sarret[1*].

[1]Department of Pharmacology-Physiology, Faculty of Medicine and Health Sciences, Institut de pharmacologie de Sherbrooke, Université de Sherbrooke, Sherbrooke, QC, Canada

[2]Department of Pharmacology and Therapeutics, McGill University, Montreal, QC, Canada

[3]Ahmanson Translational Theranostic Division, Department of Molecular and Medical Pharmacology, David Geffen School of Medicine, University of California at Los Angeles, Los Angeles, CA, USA

[‡] **Equal contribution**

[*]**Corresponding authors**

**Philippe Sarret, Ph.D.**
Dept. of Pharmacology-Physiology
Faculty of Medicine and Health Sciences
Université de Sherbrooke
3001, 12th Avenue North
Sherbrooke, Québec, J1H 5H4
Canada
Tel: (819) 821-8000, Ext: 72554
Philippe.Sarret@USherbrooke.ca

**Élie Besserer-Offroy, Ph.D.**
Dept. of Pharmacology and Therapeutics
McGill University
McIntyre Medical Sciences Building
3655 Sir William Osler Promenade
Montréal, Québec, H3G 1Y6
Canada
Tel: (514) 393-8803
Elie.Besserer@McGill.ca

**Table of contents**

## Supplementary Scheme S1. Structure of the pepducins PP-001 - PP-005

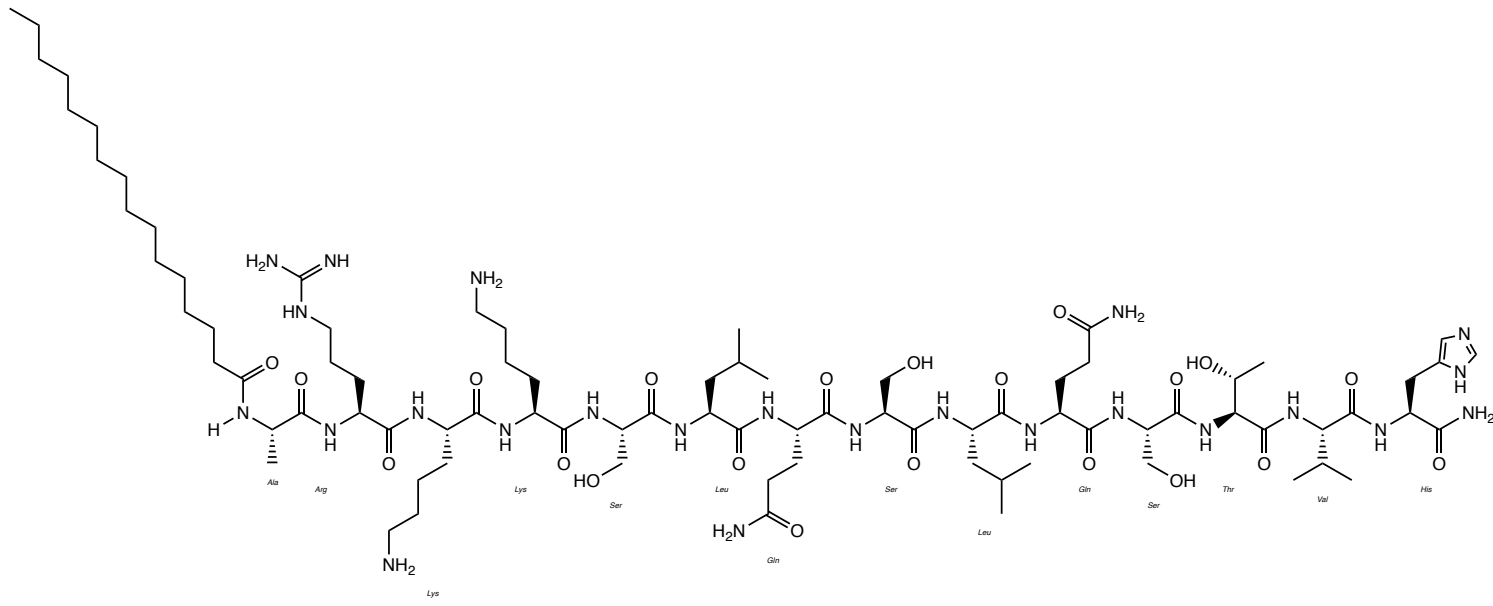

**NTS1-ICL1-PP-001**

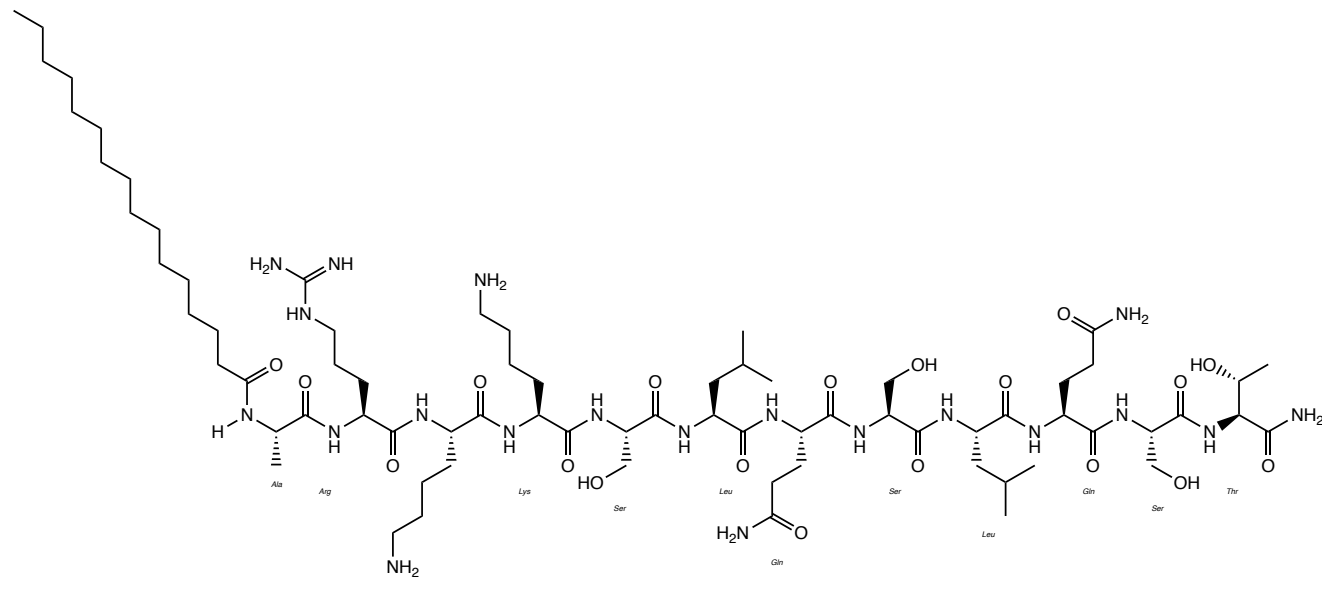

**NTS1-ICL1-PP-002**

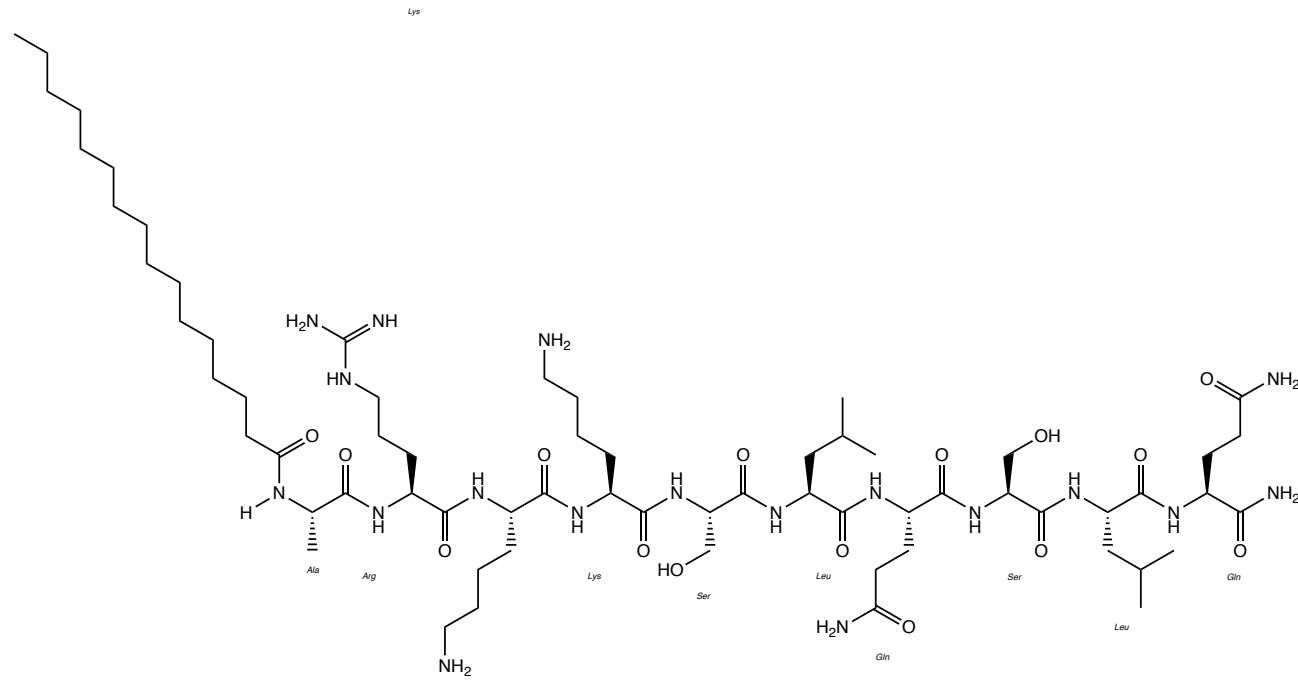

**NTS1-ICL1-PP-003**

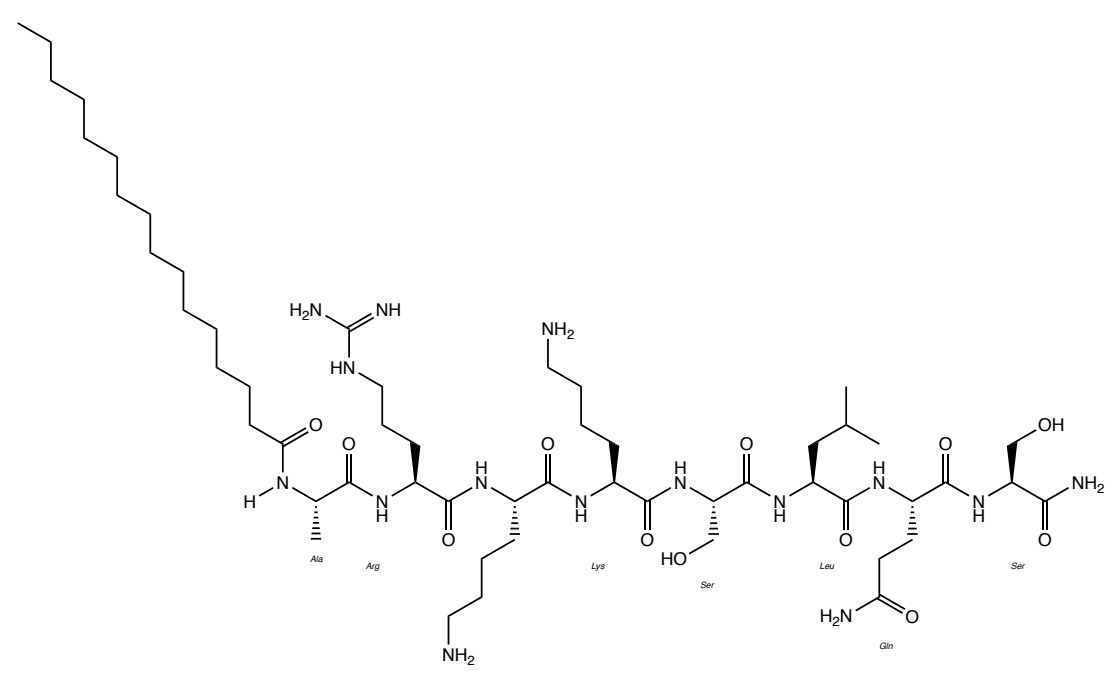

**NTS1-ICL1-PP-004**

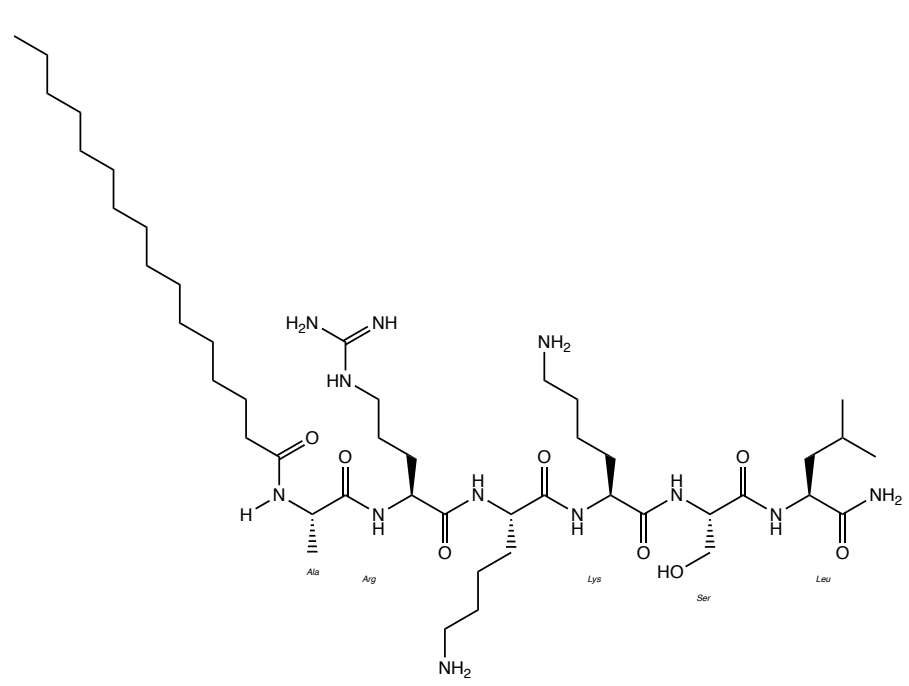

**NTS1-ICL1-PP-005**

## Supplementary Scheme S2. Structure of the control pepducins

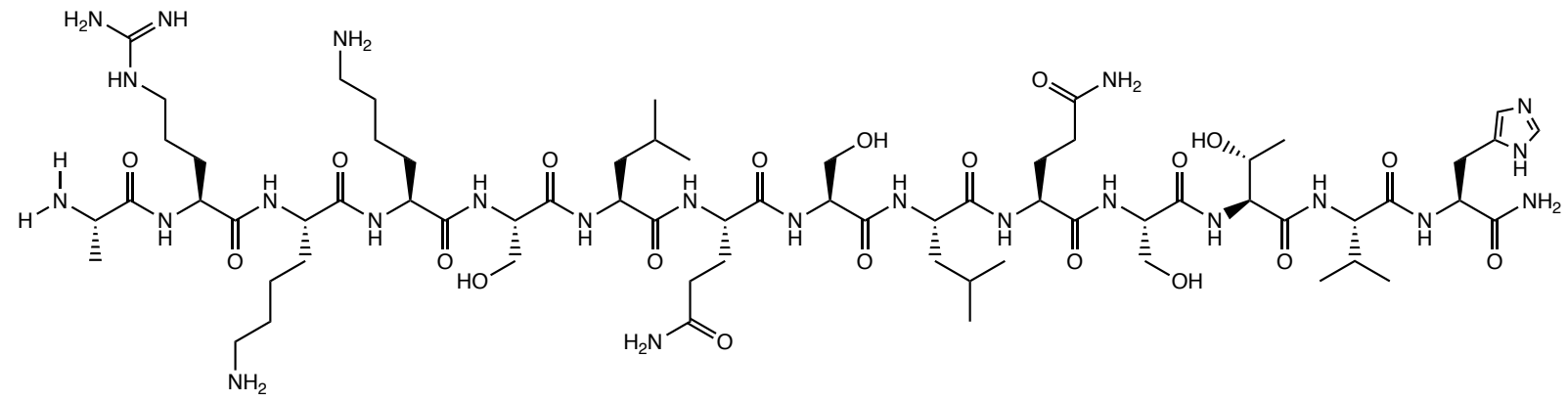

**NTS1-ICL1-NP-001**

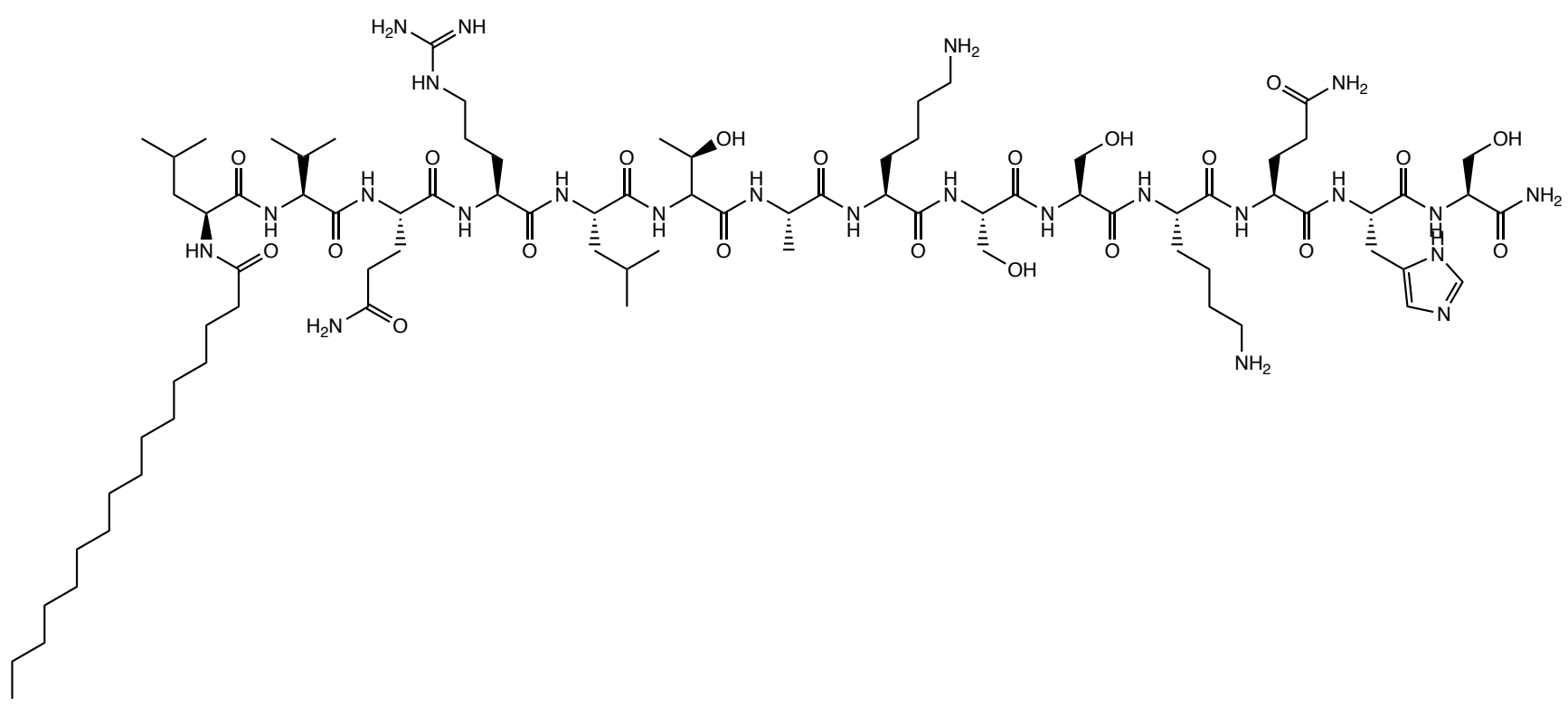

**NTS1-ICL1-PP-SCR-001**

## Supplementary Material and Methods

### *Evaluating PP-001 cytotoxicity*

Cell viability was assessed using the Cell Titer-Glo Luminescent Assay kit (Promega Corporation, Madison, WI, USA) according to manufacturer's instructions. Briefly, CHO-hNTS1 were seeded onto white opaque 96-well plates (BD Falcon, Corning, NY, USA) at 35 000 cells per well (100 μL). Twenty-four hours post-seeding, the cells were treated with the vehicle (serum-starved DMEM-F12 media) or with 10 μM PP-001, at a total volume of 100 μL per well. Additional wells received 100 μL of vehicle only (no cells) as a control. Incubation time was set for 16 hours. The plates were removed from incubator 20 min prior to assay in order to reach room temperature, and wells were treated with 100 μL of Cell Titer-Glo Reagent previously thawed to room temperature. Plates were set on a nutating mixer for 2 min and incubated at room temperature for an additional 10 min. Total luminescence counts were read on a GENios Pro plate reader (Tecan, Durham, NC, USA). Luminescence counts were normalized: the luminescence from the vehicle-treated condition was set as 100% cell viability, and the luminescence from the vehicle only condition (no cells) set as 0% cell viability. Data represent mean ± SEM, $n = 2$, tested in quadruplicate.

### *Evaluating stability of PP-001 in rat cerebrospinal fluid and plasma*

CSF was sampled from anesthetized rats directly from the cistern magna. CSF was frozen at -20˚C until use. Plasma was sampled from anesthetized rats by cardiac puncture in 4.5 mL plasma separating tubes coated with lithium heparin (from BD). Tubes were then centrifuged at 2500 rpm for 15 min at 4˚C to separate the plasma from the blood cells. Plasma was aliquoted and stored at -80˚C until use.

Stability assay was performed by adding 6 µL of PP-001 (1 mM) to 27 µL of either CSF or plasma. After mixing, the reaction was incubated at 37˚C for 60 min. Reaction was stopped over ice and 25 µL of 5% formic acid was added to separate the compound bound to plasma proteins. Then 100 µL of acetonitrile containing 0.25 mM N-N-dimethylbenzamide (as an internal standard) was added to precipitate all the proteins. The mixture was vortexed and centrifuged at 13,000 g for 30 min at 4˚C. Supernatant was filtered through 0.22 µm PTFE filters, diluted with 80 µL of $ddH_2O$ and analized by UPLC/MS (Waters UPLC system coupled with a SQ detector 2 and a PDA eλ detector, coupled to an Acquity UPLC BEH C4 column, 2.1 mm X 50 mm, 1.7 µm spherical size) using the following gradient: water + 0.1% TFA and acetonitrile (0 → 0.2 min, 5% acetonitrile; 0.2 → 1.5 min, 5% → 95%; 1.5 → 1.8 min, 95%; 1.8 → 2.0 min, 95% → 5%; 2.0 → 2.5 min, 5%). Quantification was done by determining the AUC (area under the curve) ratio of PP-001 over the AUC of the internal standard.

***Monitoring animal core-body temperature***

Body temperature was measured using a thermistor probe inserted into the rectum of adult Sprague-Dawley rats. Prior to testing, animals were individually acclimatized to manipulations and thermistor probe 5 min/day for three consecutive days. On the test day, temperature was measured before (baseline) and 75 min following i.t. injection of saline, PP-001 at 275 nmol/kg or the NTS1-selective reference compound PD149163 (1). Variations in body temperature (Δ body temp, degrees Celsius) were determined as changes from baseline for each individual animal.

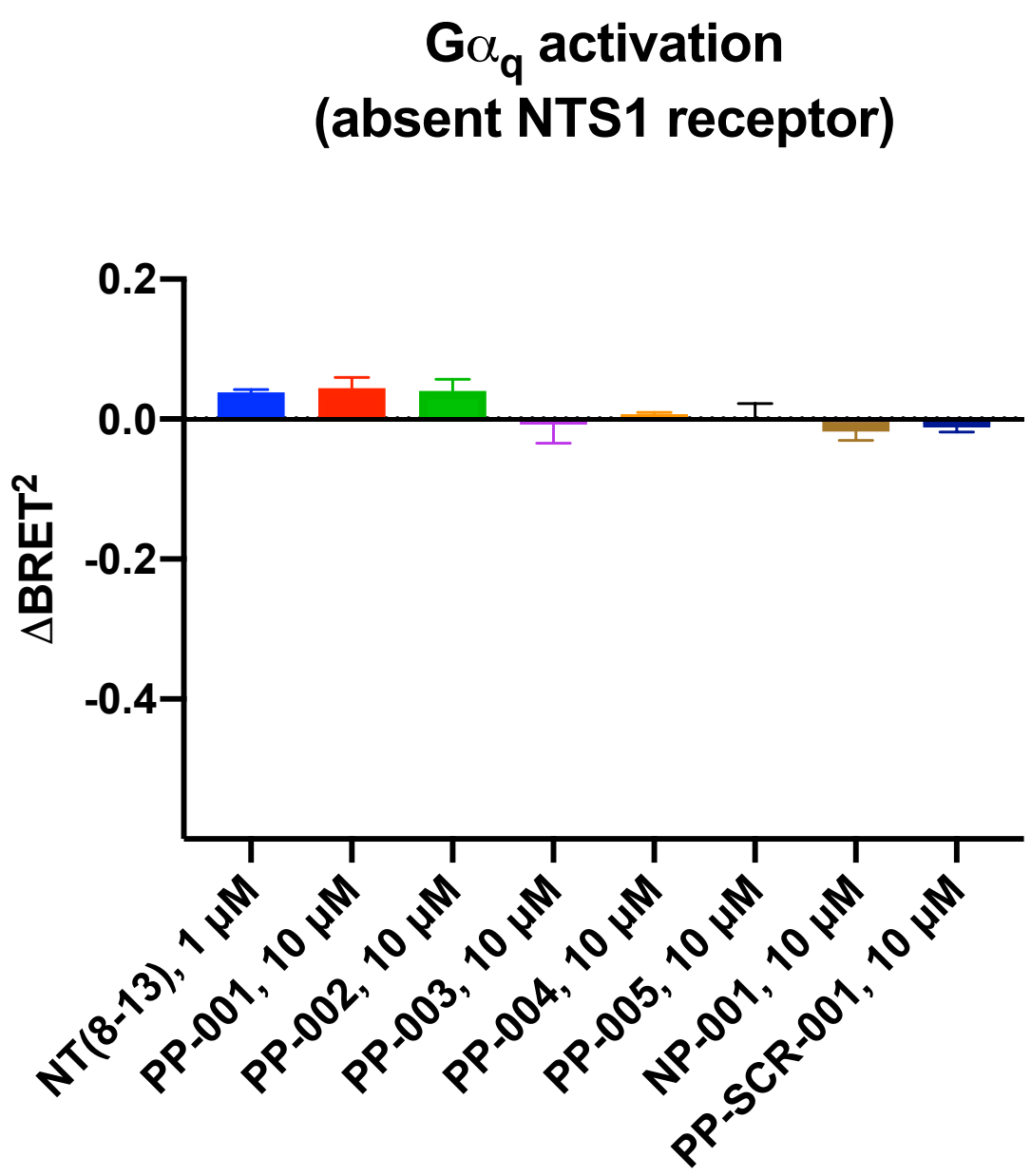

**Supplementary Figure S1. Absence of Gαq protein signaling pathway activation in mock CHO cells in response to pepducin treatment, monitored by BRET$^2$.** Gαq protein activation in CHO-K1 cells transiently transfected with the BRET$^2$-based biosensors, 15 min post-stimulation with NT(8-13) (1 μM) or with the ICL1-derived hNTS1 pepducins (10 μM). BRET$^2$ ratios at 15 min post-stimulation were subtracted from those at baseline for each condition in order to obtain the ΔBRET$^2$. In this paradigm, a decrease in BRET$^2$ corresponds to G protein activation. Data represent mean ± SEM, *n* = 2, performed in duplicate.

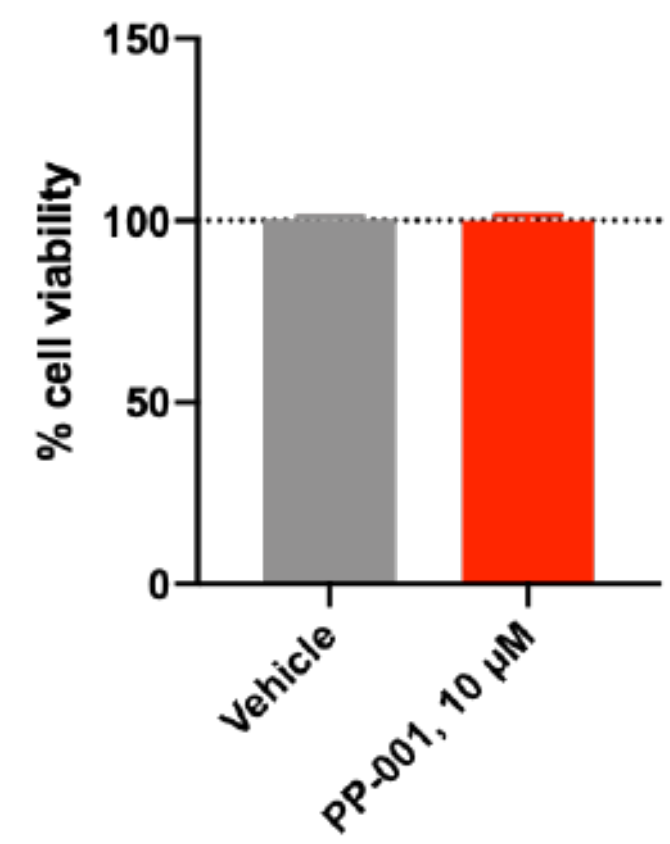

**Supplementary Figure S2. Absence of cytotoxicity in PP-001-treated cells, 16 hours post-stimulation.** Cell viability of CHO-hNTS1 cells incubated for 16 hours with the vehicle (serum-starved DMEM-F12 media) or with PP-001 at a 10 µM concentration, assessed using the Cell Titer-Glo Luminescent Assay kit by Promega. Data represent normalized luminescence counts: luminescence from the vehicle-treated condition was set as 100 % cell viability, and luminescence from the vehicle only condition (no cells) set as 0 % cell viability. Data represent mean ± SEM, *n* = 2, tested in quadruplicate.

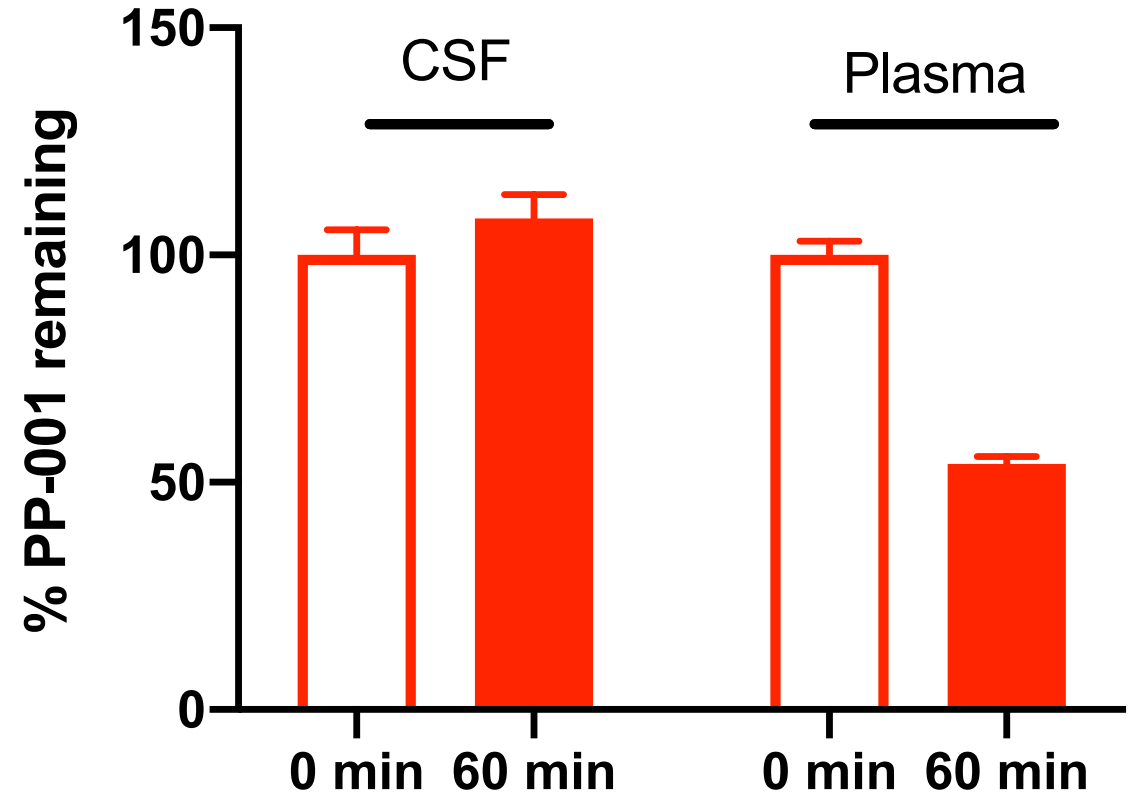

**Supplementary Figure S3**. **Stability of PP-001 pepducin in rat cerebrospinal fluid and plasma.** Quantification by UPLC/MS of PP-001 pepducin before and after a 60-min incubation period at 37°C in rat cerebrospinal fluid (CSF) and in rat plasma. Data represents mean ± SEM, *n* = 3 independent samples per condition.

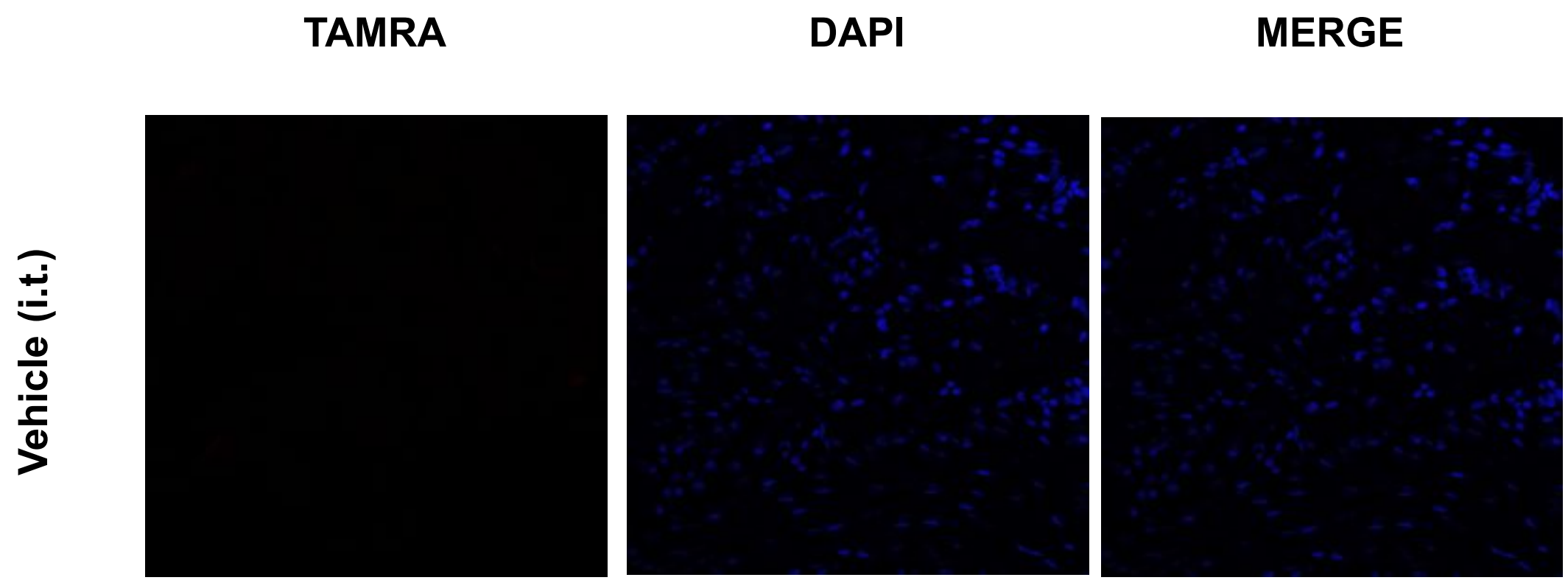

**Supplementary Figure S4. Absence of TAMRA-associated fluorescence in dorsal root ganglion neurons of vehicle-treated rats.** Confocal microscopy images of dorsal root ganglion cells extracted at the level of the L6 vertebrae from rats intrathecally injected with the vehicle (saline, 10% DMSO, 20% PEG) (*n* = 5 rats). TAMRA-associated fluorescence was visualized in the red filter channel. DAPI (blue) was used as a nuclear counterstain. Images were acquired under 40x magnification.

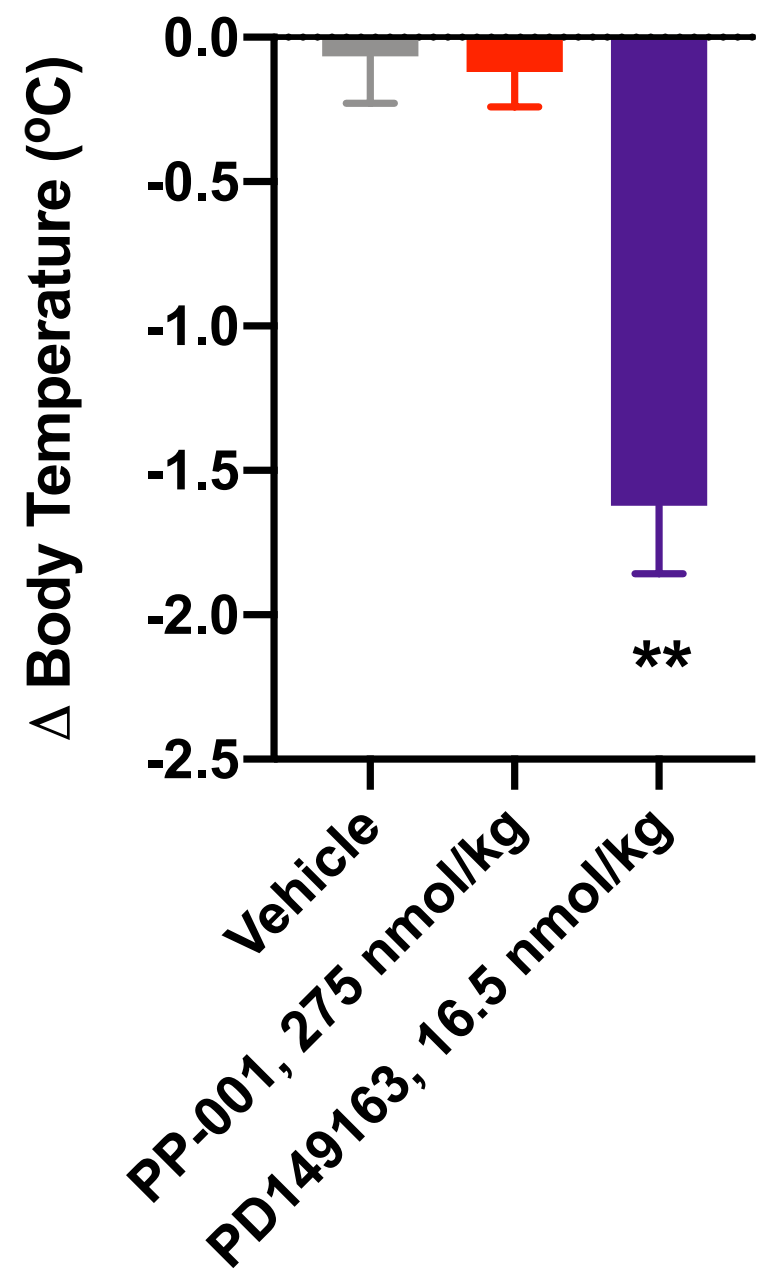

**Supplementary Figure S5. Effect of PP-001 on rat core body temperature.** Rat core body temperature changes (in degrees Celsius) induced by i.t. injection of vehicle, PP-001 (275 nmol/kg) or the NTS1-selective agonist PD149163 (16.5 nmol/kg). Baseline body temperature assessment was performed before i.t. injection. Error bars represent mean ± SEM, $n$= 6 animals per group. A Kruskal Wallis test with Dunn's correction for multiple comparisons was performed. Asterisks denote statistical differences between the vehicle and the PD149163-treated group: **, $p < 0.01$

**Supplementary Reference**